\documentclass{tlp}
\usepackage{latexsym}
\usepackage{amssymb}
\begin{document}
\title[A Backward Analysis for Constraint Logic Programs]
{\begin{tabular}{c}
A Backward Analysis for\\
Constraint Logic Programs
\end{tabular}}
\author[Andy King and Lunjin Lu]
{ANDY KING\\
University of Kent at Canterbury, Canterbury, CT2~7NF, UK. \\
email: \texttt{amk@ukc.ac.uk} \and
LUNJIN LU\\
Oakland University, Rochester, MI~48309, USA. \\
email: \texttt{l2lu@oakland.edu} }
\maketitle

\def\comment#1          {}
\def\nat        {\mathbb{N}}

\newtheorem{lemma}{Lemma}[section]
\newtheorem{proposition}{Proposition}[section]
\newtheorem{theorem}{Theorem}[section]
\newtheorem{corollary}{Corollary}[section]
\newtheorem{definition}{Definition}[section]
\newtheorem{example}{Example}[section]

\newcommand {\comments} [1]{}
\newcommand\all[1]{\mbox{${\it all}(#1)$}}
\newcommand\eprj[2]{\mbox{$\exists vars(#2).#1$}}
\def\definedas{\stackrel{def}{=}}
\def\csep{~$\|$~}

\def\land           {\wedge}
\def\true           {{\it true}}
\def\lor            {\vee}
\def\liff           {\leftrightarrow}
\def\limply         {\rightarrow}
\def\lif            {\leftarrow}
\def\lfor           {\forall}
\def\lexist         {\exists}
\def\lfrom          {\leftarrow}
\def\lto            {\rightarrow}
\def\xvec           {\bar{X}}
\def\lback          {\triangleright}
\def\mydownarrow    {{\downarrow \!}}
\def\reduce         {\rightarrow}
\def\rreduce        {\Rightarrow}
\def\rrreduce       {\Rrightarrow}
\def\separate       {\diamond}
\def\mymodels       {\unlhd}
\def\mywedge        {\otimes}
\def\myvee      {\oplus}
\def\neck           {{\, \mbox{:-} \,}}

\begin{abstract}
One recurring problem in program development is that of
understanding how to re-use code developed by a third party. In
the context of (constraint) logic programming, part of this
problem reduces to figuring out how to query a program. If the
logic program does not come with any documentation, then the
programmer is forced to either experiment with queries in an
\textit{ad hoc} fashion or trace the control-flow of the program
(backward) to infer the modes in which a predicate must be called
so as to avoid an instantiation error. This paper presents an
abstract interpretation scheme that automates the latter
technique. The analysis presented in this paper can infer moding
properties which if satisfied by the initial query, come with the
guarantee that the program and query can never generate any moding
or instantiation errors. Other applications of the analysis are
discussed. The paper explains how abstract domains with certain
computational properties (they condense) can be used to trace
control-flow backward (right-to-left) to infer useful properties
of initial queries. A correctness argument is presented and an
implementation is reported.
\end{abstract}

\section{Introduction}

The myth of the lonely logic programmer writing a program in
isolation is just that: a myth. Applications (and application
components) are usually implemented and maintained by a team. One
consequence of this is a significant proportion of the program
development effort is devoted to understanding code developed by
another. One advantage of (constraint) logic programs for software
development is that their declarative nature makes them less
opaque than, say, C++ programs. One disadvantage of logic programs
over C++ programs, however, is that the signature (argument types)
of a predicate do not completely specify how the predicate should
be invoked. In particular, a call to a predicate from an
unexpected context may generate an error if an argument of the
call is insufficiently instantiated (even if the program and query
are well-typed). This is because logic programs contain builtins
and calls to these builtins often impose moding requirements on
the query. If the program is developed by another programmer, it
may not be clear how to query a predicate so as to avoid an
instantiation error. In these circumstances, the programmer will
often resort to a trial and error tactic in their search for an
initial call mode. This can be both frustrating and tedious and,
of course, cannot guarantee coverage of all the program execution
paths. This paper presents an analysis for inferring moding
properties which, if satisfied by the initial query, ensure that
the program does not generate instantiation errors. Of course, it
does not mean that the inferred call has the form exactly intended
by the original programmer -- no analysis can do that -- the
analysis just recovers mode information. Nevertheless, this is a
useful first step in understanding the code developed by another.

The problem of inferring initial queries which do not lead to
instantiation errors is an instance of the more general problem of
deducing how to call a program so that it conforms to some desired
property, for example, calls to builtins do not error, the program
terminates, or calls to builtins behave predictably. The backward
analysis presented in this paper is designed to infer conditions
on the query which, if satisfied, guarantee that resulting
derivations satisfy a property such as one of those above.
Specifically, the analysis framework can be instantiated to solve
the following analysis problems:
\begin{itemize}

\item
Builtins and library functions can behave unpredictably when
called with infinite rational trees. For example, the query ?- X =
X + X, Y is X will not terminate in SICStus Prolog because the
arithmetic operator expects its input to be a finite tree rather
than an infinite rational tree. Moreover, the standard term
ordering of Prolog does not lift to rational trees, so the builtin
sort can behave unpredictably when sorting rational trees. These
problems (and related problems with builtins) motivate the use of
dependency analysis for tracking which terms are definitely finite
\cite{BZGH01}. The basic idea is to describe the constraint $x =
f(x_1, \ldots, x_n)$ by the Boolean function $x \Leftrightarrow
\wedge_{i = 1}^{n} x_i$ which encodes that $x$ is bound to a
finite tree iff each $x_i$ is bound to a finite tree. Although not
proposed in the context of backward analysis \cite{BZGH01}, the
framework proposed in this paper can be instantiated with a finite
tree dependency domain to infer finiteness properties on the query
which, if satisfied, guarantee that builtins are not called with
problematic arguments.

\item Termination inference is the problem of inferring initial modes for
a query that, if satisfied, ensure that a logic program
terminates. This problem generalises termination checking which
verifies program termination for a class of queries specified by a
given mode. Termination inference dates back to \cite{Mesnard96b}
but it has been recently observed \cite{GC01} that the missing
link between termination checking and termination inference is
backward analysis. A termination inference analyser is reported in
\cite{GC01} composed from two components: a standard termination
checker \cite{CT99} and the backward analysis described in this
paper. The resulting analyser is similar to the cTI analyser of
\cite{Mesnard01} -- the main difference is its design as two
existing black-box components which, according to \cite{GC01},
simplifies the formal justification and implementation.

\item
Mode analysis is useful for implementing ccp programs. In
particular \cite{DGB96} explains how various low-level
optimisations, such as returning output values in registers, can
be applied if goals can be scheduled left-to-right without
suspension. If the guards of the predicates are re-interpreted as
moding requirements, then the backward mode analysis can infer
sufficient conditions for avoiding deadlock under left-to-right
scheduling. The analysis presented in this paper thus has
applications outside program development.

\end{itemize}
\noindent To summarise, the analysis presented in this paper can
deduce properties of the call which, if satisfied, guarantee that
resulting derivations fulfill some desired property. The analysis
is unusual in that it applies lower approximation
(see~\ref{sect-lower}) as well as upper approximation
(see~\ref{sect-upper}); it is formulated in terms of a greatest
fixpoint calculation (see~\ref{sect-gfp}) as well as least
fixpoint calculation (see~\ref{sect-lfp}); the analysis also
imposes some unusual restrictions on the abstract domain
(see~\ref{sect-restrict}).

\subsection{Backward analysis}

Backward analysis has been applied extensively in functional
programming in, among other things, projection analysis
\cite{WH87}, stream strictness analysis \cite{HW89}, inverse image
analysis \cite{D91}, \textit{etc}. By reasoning about the context
of a function application, these analyses can identify
opportunities for eager evaluation that are missed by (forward)
strictness analysis as proposed by \cite{M81}. Furthermore,
backward reasoning on imperative programs dates back to the early
days of static analysis \cite{CC82}. By way of contrast, backward
analysis has been rarely applied in logic programming. One notable
exception is the demand analysis of \cite{D93}. This analysis
infers the degree of instantiation necessary for the guards of a
concurrent constraint program (ccp) to reduce. It is a local
analysis that does not consider the possible suspension of body
calls. This analysis detects those (uni-modal) predicates which
can be implemented with specialised suspension machinery. A more
elaborate backward analysis for ccp is presented by \cite{FHW00}.
This demand analysis infers how much input is necessary for a
procedure to generate a certain amount of output. This information
is useful for adding synchronisation (ask) constraints to a
procedure to delay execution and thereby increase grain size, and
yet not introduce deadlock. (Section~\ref{sect-relate} provides
more extensive and reflective review of the related work.)

\subsection{Contributions}\label{sect-contrib}

Our work is quite different. As far as we are aware, it is unique
in that it focuses on the backward analysis of (constraint) logic
programs with left-to-right scheduling. Specifically, our work
makes the following practical and theoretical contributions:
\begin{itemize}

\item it shows how to compute an initial mode of a predicate
which is safe in that if a query is at least as instantiated as
the inferred mode, the execution is guaranteed to be free from
instantiation errors. The modes inferred are often disjunctive,
sometimes surprising and, for the small predicates that we
verified by hand, appear to be optimal.

\item it specifies a practical algorithm for calculating initial modes that
is straightforward to implement in that it reduces to two
bottom-up fixpoint calculations. Furthermore, this backward
analysis problem cannot be solved with any existing abstract
interpretation machinery.

\item to the best our knowledge, it is the first time
domains that are closed under Heyting completion \cite{GS98}, or
equivalently are condensing \cite{MS93}, have been applied to
backward analysis. Put another way, our work adds credence to the
belief that condensation is an important property in the analysis
of logic programs.

\end{itemize}
The final point requires some unpacking. Condensation was
originally proposed in \cite{L91}, though arguably the simplest
statement of this property \cite{MS93} is for downward closed
domains such as $Pos$ \cite{AMSS98} and the $Pos$-like type
dependency domains \cite{CL00}. Suppose that $f : X \rightarrow X$
is an abstract operation on a downward closed domain $X$ equipped
with an operation $\wedge$ that mimics unification or constraint
solving. $X$ is condensing iff $x \wedge f(y) = f(x \wedge y)$ for
all $x, y \in X$. Hence, if $X$ is condensing, $x \wedge f(true) =
f(x)$ where $true$ represents the weakest abstract constraint.
More exactly, if $f(true)$ represents the result of the
goal-independent analysis, and $f(x)$ the result of the
goal-dependent one with an initial constraint $x$, then the
equivalence $f(x) = x \wedge f(true)$ enables goal-dependent
analysis to be performed in a goal-independent way without loss of
precision. This, in turn, can simplify the implementation of an
analyser \cite{AMSS98}. Because of this, domain refinement
machinery has been devised to enrich a domain with new elements to
obtain the desired condensing property \cite{GS98}. It turns out
that it is always possible to systematically design a condensing
domain for a given downward closed property
\cite{GS98}[Theorem~8.2] by applying Heyting completion.
Conversely, under some reasonable hypotheses, all condensing
domains can be reconstructed by Heyting completion
\cite{GS98}[Theorem~8.3]. One consequence of this is that
condensing domains come equipped with a (pseudo-complement)
operator and this turns out to be an operation that is important
in backward analysis. To summarise, machinery has been developed
to synthesise condensing domains and condensing domains provide
operations suitable for backward analysis.

\subsection{Organisation of the paper}

The rest of the paper is structured as follows.
Section~\ref{sect-informal} introduces the key ideas of the paper
in an informal way through a worked example.
Section~\ref{sect-prelims} introduces the necessary preliminaries
for the formal sections that follow. Section~\ref{sect-assert}
presents an operational semantics for constraint logic programs
with assertions in which the set of program states is augmented by
a special error state. Section~\ref{sect-collect} develops a
semantics which computes those initial states that cannot lead to
the error state. The semantics defines a framework for backward
analysis and formally argues correctness.
Section~\ref{sect-implement} describes an instantiation of the
framework for mode analysis. Section~\ref{sect-relate} reviews the
related work and section~\ref{sect-conclude} concludes. Much of
the formal machinery is borrowed directly from \cite{GDL95,GS98}
and in particular the reader is referred to \cite{GDL95} for
proofs of the semantic results stated in
section~\ref{sect-prelims} (albeit presented in a slightly
different form). To aid continuity in the paper, the remaining
proofs are relegated to appendix~\ref{sect-appendix}.

\section{Worked example}\label{sect-informal}

\subsection{Basic components}

This section informally presents an abstract interpretation scheme
which infers how to query a given predicate so as to avoid
run-time moding errors. In other words, the analysis deduces
moding properties of the call that, if satisfied, guarantee that
resulting derivations cannot encounter an instantiation error. To
illustrate, consider the Quicksort program listed in the left
column of figure~\ref{fig-qsort-three}. This is the first
ingredient of the analysis: the input program. The second
ingredient is an abstract domain which, in this case, is $Pos$.
$Pos$ is the domain of positive Boolean functions, that is, the
set of functions $f : \{0,1\}^n \to \{0,1\}$ such that \mbox{$f(1,
\ldots, 1) = 1$}. Hence $x \vee y \in Pos$ since $1 \vee 1 = 1$
but $\neg x \not\in Pos$ since $\neg 1 = 0$. $Pos$ is augmented
with the bottom element $0$ with $1$ being the top element. The
domain is ordered by entailment $\models$ and, in this example,
will be used to represent grounding dependencies.

$Pos$ comes equipped with the logical operations: conjunction
$\wedge$, disjunction $\vee$, implication $\Rightarrow$ (and thus
bi-implication $\Leftrightarrow$). Conjunction is used to conjoin
the information from different body atoms, while disjunction is
used to combine the information from different clauses.
Conjunction and disjunction, in turn, enable two projection
operators to be defined: $\exists_x(f) = f[x \mapsto 0] \vee f[x
\mapsto 1]$ and $\forall_x(f) = f'$ if $f' \in Pos$ otherwise
$\forall_x(f) = 0$ where $f' = f[x \mapsto 0] \wedge f[x \mapsto
1]$. Note that although $f[x \mapsto 0] \vee f[x \mapsto 1] \in
Pos$ for all $f \in Pos$ it does not follow that $f[x \mapsto 0]
\wedge f[x \mapsto 1] \in Pos$ for all $f \in Pos$. Indeed, $(x
\Leftarrow y)[x \mapsto 0] \wedge (x \Leftarrow y)[x \mapsto 1] =
\neg y$. Both operators are used to project out the body variables
that are not in the head of a clause. Specifically, these
operators eliminate the variable $x$ from the formula $f$. They
are dual in the sense that $\forall_x(f) \models f \models
\exists_x(f)$. These are the basic components of the analysis.

\subsection{Normalisation and abstraction}

The analysis components are assembled in two steps. The first is a
bottom-up analysis for success patterns, that is, a bottom-up
analysis which infers the groundness dependencies which are known
to be created by each predicate regardless of the calling pattern.
This step is a least fixpoint (lfp) calculation. The second step
is a bottom-up analysis for input modes (the objective of the
analysis). This step is a greatest fixpoint (gfp) computation. To
simplify both steps, the program is put into a form in which the
arguments of head and body atoms are distinct variables. This
gives the normalised program listed in the centre column of
figure~\ref{fig-qsort-three}. This program is then abstracted by
replacing each Herbrand constraint \mbox{$x = f(x_1, \ldots,
x_n)$} with a formula  \mbox{$x \Leftrightarrow \wedge_{i=1}^{n}
x_i$} that describes its grounding dependency. This gives the
abstract program listed in the right column of
figure~\ref{fig-qsort-three}. The 
formula $1$ in the assertion represents $true$
whereas the
formulae $g_i$ that appear in
the abstract program are as follows:
\[ \begin{array}{r@{\; = \;}l}
g_1 & t_1 \wedge (t_2 \Leftrightarrow s) \\
g_2 & t_1 \Leftrightarrow (m \wedge xs) \wedge t_3 \Leftrightarrow (m \wedge r) \\
g_3 & t_1 \wedge t_2 \wedge t_3
\end{array}
\quad
\begin{array}{r@{\; = \;}l}
g_4 & t_1 \Leftrightarrow (x \wedge xs) \wedge t_2 \Leftrightarrow (x \wedge l) \\
g_5 & t_1 \Leftrightarrow (x \wedge xs) \wedge t_2 \Leftrightarrow (x \wedge h) \\
g_6 & m \wedge x
\end{array} \]
Builtins that occur in the source, such as the tests $\mathtt{=<}$
and $\mathtt{>}$, are handled by augmenting the abstract program
with fresh predicates, $\mathtt{=<'}$ and $\mathtt{>'}$, which
express the grounding behaviour of the builtins. The $\separate$
symbol separates an assertion (the required mode) from another
$Pos$ formula describing the grounding behaviour of a successful
call to the builtin (the success mode). For example, the formula
$g_6$ left of $\separate$ in the $\mathtt{=<'}$ clause asserts
that the $\mathtt{=<}$ test will error if its first two arguments
are not ground, whereas the $g_6$ right of $\separate$ describes
the state that holds if the test succeeds. These formulae do not
coincide for all builtins (see Table~\ref{table-abstract}). For
quicksort, the only non-trivial assertions arise from builtins.
This would change if the programmer introduced assertions for
verification \cite{PBH00a}.

\begin{figure}
\begin{center}
\begin{tabular}{@{}p{1.7in}|p{1.5in}|p{1.5in}@{}}
\vspace{-0.3in} \texttt{\begin{tabbing}
xxx \= \kill \\
qs([], $s$, $s$). \\
qs([$m | xs$], $s$, $t$) :- \\
\> pt($xs, m, l, h$), \\
\> qs($l$, $s$, [$m|r$]), \\
\> qs($h$, $r$, $t$). \\
\\
pt([], \_, [], []). \\
pt([$x|xs$], $m$, [$x|l$], $h$) :- \\
\> $m$ =< $x$, \\
\> pt($xs, m, l, h$). \\
pt([$x|xs$], $m$, $l$, [$x|h$]) :- \\
\> $m$ > $x$, \\
\> pt($xs, m, l, h$).
\end{tabbing}}
\vspace{-0.2in} & \vspace{-0.3in} \texttt{\begin{tabbing}
xxx \= \kill \\
qs($t_1, s, t_2$) :- \\
\> $t_1$ = [], $t_2$ = $s$. \\
qs($t_1, s, t$) :- \\
\> $t_1$ = [$m | xs$], \\
\> $t_3$ = [$m | r$], \\
\> pt($xs, m, l, h$), \\
\> qs($l, s, t_3$), \\
\> qs($h, r, t$). \\
\\
pt($t_1, \_, t_2, t_3$) :- \\
\> $t_1$ = [], \\
\> $t_2$ = [], $t_3$ = []. \\
pt($t_1, m, t_2, h$) :- \\
\> $t_1$ = [$x|xs$], \\
\> $t_2$ = [$x|l$], \\
\> $m$ =< $x$ \\
\> pt($xs, m, l, h$). \\
pt($t_1, m, l, t_2$) :- \\
\> $t_1$ = [$x|xs$], \\
\> $t_2$ = [$x|h$], \\
\> $m$ > $x$, \\
\> pt($xs, m, l, h$).
\end{tabbing}}
\vspace{-0.2in} & \vspace{-0.3in} \texttt{\begin{tabbing}
xxx \= \kill \\
qs($t_1, s, t_2$) :- \\
\> $1 \separate g_1$. \\
qs($t_1, s, t$) :- \\
\> $1 \separate g_2$, \\
\> pt($xs, m, l, h$), \\
\> qs($l, s, t_3$), \\
\> qs($h, r, t$). \\
\\
pt($t_1, \_, t_2, t_3$) :- \\
\> $1 \separate g_3$. \\
pt($t_1, m, t_2, h$) :- \\
\> $1 \separate g_4$, \\
\> =<'($m$, $x$), \\
\> pt($xs, m, l, h$). \\
pt($t_1, m, l, t_2$) :- \\
\> $1 \separate g_5$, \\
\> >'($m$, $x$), \\
\> pt($xs, m, l, h$). \\
\\
=<'($m$, $x$) :- $g_6 \separate g_6$. \\
\\
>'($m$, $x$) :- $g_6 \separate g_6$.
\end{tabbing}} \vspace{-0.2in}
\end{tabular}
\end{center}
\caption{Quicksort: raw, normalised and
abstracted}\label{fig-qsort-three}
\end{figure}

\subsection{Least fixpoint calculation}\label{sect-lfp}

An iterative algorithm is used to compute the lfp and thereby
characterise the success patterns of the program. A success
pattern is a pair consisting of an atom with distinct variables
for arguments paired with a $Pos$ formula over those variables.
Renaming and equality of formulae induce an equivalence between
success patterns which is needed to detect the fixpoint. The
patterns $\langle p(u, w, v), u \wedge (w \Leftrightarrow v)
\rangle$ and $\langle p(x_1, x_2, x_3), (x_3 \Leftrightarrow x_2)
\wedge x_1 \rangle$, for example, are considered to be identical:
both express the same inter-argument groundness dependencies. Each
iteration produces a set of success patterns: at most one pair for
each predicate in the program.

\subsubsection{Upper approximation of success patterns}\label{sect-upper}

A success pattern records an inter-argument groundness dependency
that describes the binding effects of executing a predicate. If
$\langle p(\vec{x}), f \rangle$ correctly describes the predicate
$p$, and $g$ holds whenever $f$ holds, then $\langle p(\vec{x}), g
\rangle$ also correctly describes $p$. Success patterns can thus
be approximated from \textit{above} without compromising
correctness.

Iteration is performed in a bottom-up fashion and commences with
$F_0 = \emptyset$. $F_{j+1}$ is computed from $F_j$ by considering
each clause \mbox{$p(\vec{x}) \leftarrow d \separate f,
p_{1}(\vec{x}_1), \ldots, p_{n}(\vec{x}_n)$} in turn. Initially
$F_{j+1} = \emptyset$. The success pattern formulae $f_i$ for the
$n$ body atoms are conjoined with $f$ to obtain \mbox{$g = f
\wedge \wedge_{i=1}^{n} f_i$}. Variables not present in
$p(\vec{x})$, $Y$ say, are then eliminated from $g$ by computing
\mbox{$g' = \exists_{Y}(g)$} (weakening $g$) where $\exists_{\{
y_1 \ldots y_n \}}(g) = \exists_{y_1}(\ldots\exists_{y_n}(g))$.
Weakening $g$ does not compromise correctness because success
patterns can be safety approximated from above.

\subsubsection{Weakening upper approximations}

If $F_{j+1}$ already contains a pattern of the form $\langle
p(\vec{x}), g'' \rangle$, then this pattern is replaced with
$\langle p(\vec{x}), g' \vee g'' \rangle$, otherwise $F_{j+1}$ is
revised to include $\langle p(\vec{x}), g' \rangle$. Thus the
success patterns become progressively weaker on each iteration.
Again, correctness is preserved because success patterns can be
safety approximated from above.

\subsubsection{Least fixpoint calculation for Quicksort}

For brevity, let $\vec{u} = \langle x_1, x_2 \rangle$, $\vec{v} =
\langle x_1, x_2, x_3 \rangle$ and $\vec{w} = \langle x_1, x_2,
x_3, x_4 \rangle$. Then the lfp for the abstracted Quicksort
program is obtained (and checked) in the following 3 iterations:
\[
F_1 = \left\{
\begin{array}{@{}r@{\,}l@{}}
\langle \mathtt{qs}(\vec{v}), & x_1 \wedge (x_2 \Leftrightarrow x_3) \rangle \\
\langle \mathtt{pt}(\vec{w}), & x_1 \wedge x_3 \wedge x_4 \rangle \\
\langle \mathtt{=<'}(\vec{u}), & x_1 \wedge x_2 \rangle \\
\langle \mathtt{>'}(\vec{u}), & x_1 \wedge x_2 \rangle
\end{array} \right\}
\quad F_2 = \left\{
\begin{array}{@{}r@{\,}l@{}}
\langle \mathtt{qs}(\vec{v}), & x_2 \Leftrightarrow (x_1 \wedge x_3) \rangle \\
\langle \mathtt{pt}(\vec{w}), & x_1 \wedge x_3 \wedge x_4 \rangle \\
\langle \mathtt{=<'}(\vec{u}), & x_1 \wedge x_2 \rangle \\
\langle \mathtt{>'}(\vec{u}), & x_1 \wedge x_2 \rangle
\end{array} \right\}
\]
Finally, $F_3 = F_2$. The space of success patterns forms a
complete lattice which ensures that a lfp (a most precision
solution) exists. The iterative process will always terminate
since the space is finite and hence the number of times each
success pattern can be updated is also finite. Moreover, it will
converge onto the lfp since iteration commences with the bottom
element $F_0 = \emptyset$.

Observe that $F_2$, the lfp, faithfully describes the grounding
behaviour of quicksort: a $\mathtt{qs}$ goal will ground its
second argument if it is called with its first and third arguments
already ground and \textit{vice versa}. Note that assertions are
not considered in the lfp calculation.

\subsection{Greatest fixpoint calculation}\label{sect-gfp}

A bottom-up strategy is used to compute a gfp and thereby
characterise the safe call patterns of the program. A safe call
pattern describes queries that do not violate the assertions. A
call pattern has the same form as a success pattern (so there is
one call pattern per predicate rather than one per clause). One
starts with assuming no call causes an error and then
checks this assumption by reasoning backwards over all clauses. If
an assertion is violated, the set of safe call patterns for
the involved predicate is strengthened (made smaller), and the whole
process is repeated until the assumptions turn out to be valid (the
gfp is reached).

\subsubsection{Lower approximation of safe call patterns}\label{sect-lower}

Iteration commences with $D_0 = \{ \langle p(\vec{x}), 1 \rangle
\mid p \in \Pi \}$ where $\Pi$ is the set of predicate symbols
occurring in the program. An iterative algorithm incrementally
\textit{strengthens} the call pattern formulae until they only
describe queries which lead to computations that satisfy the
assertions. Note that call patterns describe a subset (rather than
a superset) of those queries which are safe. Call patterns are
thus lower approximations in contrast to success patterns which
are upper approximations. Put another way, if $\langle p(\vec{x}),
g \rangle$ correctly describes some safe call patterns of $p$, and
$g$ holds whenever $f$ holds, then $\langle p(\vec{x}), f \rangle$
also correctly describes some safe call patterns of $p$. Call
patterns can thus be approximated from \textit{below} without
compromising correctness (but not from above).

$D_{k+1}$ is computed from $D_k$ by considering each
\mbox{$p(\vec{x}) \leftarrow d \separate f, p_{1}(\vec{x}_1),
\ldots, p_{n}(\vec{x}_n)$} in turn and calculating a formula that
characterises its safe calling modes. Initially set $D_{k+1} =
D_k$. A safe calling mode is calculated by propagating moding
requirements right-to-left by repeated application of the logical
operator $\Rightarrow$. More exactly, let $f_i$ denote the success
pattern formula for $p_{i}(\vec{x}_i)$ in the previously computed
lfp and let $d_i$ denote the call pattern formula for
$p_{i}(\vec{x}_i)$ in $D_k$. Set $e_{n+1} = 1$ and then compute
$e_{i} = d_i \wedge (f_i \Rightarrow e_{i+1})$ for $1 \leq i \leq
n$. Each $e_i$ describes a safe calling mode for the compound goal
\mbox{$p_{i}(\vec{x}_i), \ldots, p_{n}(\vec{x}_n)$}.


\subsubsection{Intuition and explanation}

The intuition behind the symbolism is that $d_i$ represents the
demand that is already known for $p_{i}(\vec{x}_i)$ not to error
whereas $e_i$ is $d_i$ possibly strengthened with extra demand
so as to ensure that the sub-goal \mbox{$p_{i+1}(\vec{x}_{i+1}),
\ldots, p_{n}(\vec{x}_n)$} also does not error when
executed immediately after $p_{i}(\vec{x}_{i})$. Put another way,
anything larger than $d_i$ may
possibly cause an error when executing $p_{i}(\vec{x}_i)$
and anything larger than $e_{i}$ may
possibly cause an error when executing \mbox{$p_{i}(\vec{x}_{i}),
\ldots, p_{n}(\vec{x}_n)$}.

The basic inductive step in the analysis is to compute an $e_i$ which ensures
that
\mbox{$p_{i}(\vec{x}_{i}), \ldots, p_{n}(\vec{x}_n)$}
does not error, given $d_i$ and $e_{i+1}$ which
respectively
ensure that
$p_{i}(\vec{x}_{i})$ and
\mbox{$p_{i+1}(\vec{x}_{i+1}), \ldots, p_{n}(\vec{x}_n)$}
do not error.
This step translates a demand
after the call to $p_{i}(\vec{x}_i)$ into a demand before the call
to $p_{i}(\vec{x}_i)$.
The tactic is to set $e_{n+1} = 1$ and then
compute $e_{i} =
d_{i} \wedge (f_{i} \Rightarrow e_{i+1})$ for $i \leq n$.
This tactic is best explained by unfolding the definitions of
$e_n$, then $e_{n-1}$, then $e_{n-2}$, and so on. This reverse
ordering reflects the order in which the $e_i$ are computed; the
$e_i$ are computed whilst walking backward across the clause.
Any calling mode is safe for the empty
goal and hence $e_{n+1} = 1$.
Note that $e_{n} = d_{n} \wedge (f_{n} \Rightarrow e_{n+1}) =
d_{n} \wedge (\neg f_{n} \vee 1) = d_{n}$. Hence $e_{n}$
represents a safe calling mode for the goal
\mbox{$p_{n}(\vec{x}_n)$}. 

Observe that $e_i$
should not be larger than $d_i$, otherwise an error may occur while
executing $p_{i}(\vec{x}_i)$. Observe too that if
\mbox{$p_{i}(\vec{x}_{i}),\ldots,p_{n}(\vec{x}_{n})$} is called
with a mode described by $d_{i}$, then
\mbox{$p_{i+1}(\vec{x}_{i+1}),\ldots,p_{n}(\vec{x}_{n})$} is
called with a mode described by \mbox{$(d_{i} \wedge f_{i})$} since
$f_{i}$ describes the success patterns of
\mbox{$p_{i}(\vec{x}_{i})$}. 
The mode $(d_i \wedge f_i)$ may satisfy the $e_{i+1}$ demand.
If it does not, then the minimal extra demand is added
to $(d_i \wedge f_i)$ so as to satisfy $e_{i+1}$.
This minimal extra demand is
\mbox{$((d_i \wedge f_i) \Rightarrow e_{i+1})$}
-- the \textit{weakest} mode that, in conjunction
with $(d_i \wedge f_i)$, ensures that $e_{i+1}$ holds. Put another
way, $((d_{i} \wedge f_{i}) \Rightarrow e_{i+1}) = \vee \{ f \in
Pos \mid (d_{i} \wedge f_{i}) \wedge f \models e_{i+1} \}$.

Combining the
requirements to satisfy 
$p_{i}(\vec{x}_i)$ and then 
\mbox{$p_{i+1}(\vec{x}_{i+1}),\ldots,p_{n}(\vec{x}_{n})$},
gives
$e_i$ = \mbox{$d_i \wedge ((d_i \wedge f_i) \Rightarrow e_{i+1})$}
which reduces to
$e_i = d_i \wedge (f_i \Rightarrow e_{i+1})$ and
corresponds to the tactic used in the basic inductive step.

\subsubsection{Pseudo-complement}

This step of calculating the \textit{weakest} mode that when
conjoined with $d_i \wedge f_i$ implies $e_{i+1}$, is the very
heart of the analysis. Setting $e_i = 0$ would trivially achieve
safety, but $e_i$ should be as weak as possible to maximise the
class of safe queries inferred. For $Pos$, computing the weakest
$e_i$ reduces to applying the $\Rightarrow$ operator, but more
generally, this step amounts to applying the pseudo-complement
operator. The pseudo-complement operator (if it exists for a given
abstract domain) takes, as input, two abstractions and returns, as
output, the \textit{weakest} abstraction whose conjunction with
the first input abstraction is at least as strong as the second
input abstraction. If the domain did not possess a
pseudo-complement, then there is not always a \textit{unique}
weakest abstraction (whose conjunction with one given abstraction
is at least as strong as another given abstraction).

To see this, consider the domain $Def$ \cite{AMSS98} which does
not possess a pseudo-complement. $Def$ is the sub-class of $Pos$
that is definite \cite{AMSS98}. This means that $Def$ has the
special property that each of its Boolean functions can be
expressed as a (possibly empty) conjunction of propositional Horn
clauses. 
As with $Pos$, $Def$ is assumed
to be augmented with the bottom element $0$.
$Def$ can thus represent the grounding dependencies
\mbox{$x \wedge y$}, $x$, \mbox{$x \Leftrightarrow y$}, $y$,
\mbox{$x \Leftarrow y$}, \mbox{$x \Rightarrow y$}, $0$ and $1$ but
\textit{not} \mbox{$x \vee y$}. 
Suppose that $d_i \wedge f_i = (x
\Leftrightarrow y)$ and $e_{i+1} = (x \wedge y)$. Then conjoining
$x$ with $d_i \wedge f_i$ would be at least as strong as $e_{i+1}$
and symmetrically conjoining $y$ with $d_i \wedge f_i$ would be at
least as strong as $e_{i+1}$. However, $Def$ does not contain a
Boolean function strictly weaker than both $x$ and $y$, namely $x
\vee y$, whose conjunction with $d_i \wedge f_i$ is at least as
strong as $e_{i+1}$. Thus setting $e_i = x$ or $e_i = y$ would be
safe but setting $e_i = (x \vee y)$ is prohibited because $x \vee
y$ falls outside $Def$. Moreover, setting $e_i = 0$ would loose
an unacceptable degree of precision. A choice would thus have to
be made between setting $e_i = x$ and $e_i = y$ in some arbitrary
fashion, so there would be no clear tactic for maximising
precision.

Returning to the compound goal
\mbox{$p_{i}(\vec{x}_{i}),\ldots,p_{n}(\vec{x}_{n})$}, a call
described by the mode $d_i \wedge ((d_{i} \wedge f_{i})
\Rightarrow e_{i+1})$ is thus sufficient to ensure that neither
\mbox{$p_{i}(\vec{x}_{i})$} nor the sub-goal
\mbox{$p_{i+1}(\vec{x}_{i+1}),\ldots,p_{n}(\vec{x}_{n})$} error.
Since $d_i \wedge ((d_{i} \wedge f_{i}) \Rightarrow e_{i+1})$ =
$d_i \wedge (f_{i} \Rightarrow e_{i+1})$ = $e_i$ it follows that
\mbox{$p_{i}(\vec{x}_{i}),\ldots,p_{n}(\vec{x}_{n})$} will not
error if its call is described by $e_i$. In particular, it follows
that $e_1$ describes a safe calling mode for the body atoms of the
clause \mbox{$p(\vec{x}) \leftarrow d \separate f,
p_{1}(\vec{x}_1), \ldots, p_{n}(\vec{x}_n)$}.

The next step is to calculate $g = d \wedge (f \Rightarrow e_1)$.
The abstraction $f$ describes the grounding behaviour of the
Herbrand constraint added to the store prior to executing the body
atoms. Thus $(f \Rightarrow e_1)$ describes the \textit{weakest}
mode that, in conjunction with $f$, ensures that $e_{1}$ holds,
and hence the body atoms are called safely. Hence $d \wedge (f
\Rightarrow e_1)$ represents the weakest demand that both
satisfies the body atoms and the assertion $d$. One subtlety which
relates to the abstraction process, is that $d$ is required to be
a lower-approximation of the assertion whereas $f$ is required to
be an upper-approximation of the constraint. Put another way, if
the mode $d$ describes the binding on the store, then the
(concrete) assertion is satisfied, whereas if the (concrete)
constraint is added to the store, then the store is described by
the mode $f$. Table~1 details how to abstract various builtins for
groundness for a declarative subset of ISO~Prolog.

\subsubsection{Strengthening lower approximations}

Variables not present in $p(\vec{x})$, $Y$ say, are then
eliminated by \mbox{$g' = \forall_{Y}(g)$} (\textit{strengthening}
$g$) where $\forall_{\{ y_1 \ldots y_n \}}(g) =
\forall_{y_1}(\ldots\forall_{y_n}(g))$. A safe calling mode for
this particular clause is then given by $g'$. Eliminating
variables from $g$ by strengthening $g$ is unusual and initially
appears strange. Recall, however, that call patterns can be
approximated from below without compromising correctness (but not
from above). In particular the standard projection tactic of
computing $\exists_{\{ y_1 \ldots y_n \}}(g)$ would result in an
upper approximation of $g$ that possibly describes a
\textit{larger} set of concrete call patterns which would be
incorrect. The direction of approximation thus dictates that
eliminating the variables $Y$ from $g$ must strengthen $g$.
Indeed, $g$ holds whenever $\forall_{y_i}(g)$ holds and therefore
$g$ holds whenever $\forall_{\{ y_1 \ldots y_n \}}(g)$ holds as
required.

$D_{k+1}$ will contain a call pattern $\langle p(\vec{x}), g''
\rangle$ and, assuming $g' \wedge g'' \neq g''$, this is updated
with $\langle p(\vec{x}), g' \wedge g'' \rangle$. Thus the call
patterns become progressively stronger on each iteration.
Correctness is preserved because call patterns can be safely
approximated from below. The space of call patterns forms a
complete lattice which ensures that a gfp exists. In fact, because
call patterns are approximated from below, the gfp is the most
precise solution, and therefore the desired solution. (This
contrasts to the norm in logic program analysis where
approximation is from above and the lfp is the most precise
solution). Moreover, since the space of call patterns is finite,
termination is assured. In fact, the scheme will converge onto the
gfp since iteration commences with the top element $D_0 = \{
\langle p(\vec{x}), 1 \rangle \mid p \in \Pi \}$.

\subsubsection{Greatest fixpoint calculation for Quicksort}

Under this procedure Quicksort generates the following $D_k$
sequence:
\[
\begin{array}{@{}r@{\;\;}l@{}}
D_0 = \left\{
\begin{array}{@{}r@{\,}l@{}}
\langle \mathtt{qs}(\vec{v}), & 1 \rangle \\
\langle \mathtt{pt}(\vec{w}), & 1 \rangle \\
\langle \mathtt{=<'}(\vec{u}), & 1 \rangle \\
\langle \mathtt{>'}(\vec{u}), & 1 \rangle
\end{array} \right\}
& D_1 = \left\{
\begin{array}{@{}r@{\,}l@{}}
\langle \mathtt{qs}(\vec{v}), & 1 \rangle \\
\langle \mathtt{pt}(\vec{w}), & 1 \rangle \\
\langle \mathtt{=<'}(\vec{u}), & x_1 \wedge x_2 \rangle \\
\langle \mathtt{>'}(\vec{u}), & x_1 \wedge x_2 \rangle
\end{array} \right\} \\
D_2 = \left\{
\begin{array}{@{}r@{\,}l@{}}
\langle \mathtt{qs}(\vec{v}), & 1 \rangle \\
\langle \mathtt{pt}(\vec{w}), & x_2 \wedge (x_1 \vee (x_3 \wedge x_4)) \rangle \\
\langle \mathtt{=<'}(\vec{u}), & x_1 \wedge x_2 \rangle \\
\langle \mathtt{>'}(\vec{u}), & x_1 \wedge x_2 \rangle
\end{array} \right\}
& D_3 = \left\{
\begin{array}{@{}r@{\,}l@{}}
\langle \mathtt{qs}(\vec{v}), & x_1 \rangle \\
\langle \mathtt{pt}(\vec{w}), & x_2 \wedge (x_1 \vee (x_3 \wedge x_4)) \rangle \\
\langle \mathtt{=<'}(\vec{u}), & x_1 \wedge x_2 \rangle \\
\langle \mathtt{>'}(\vec{u}), & x_1 \wedge x_2 \rangle
\end{array} \right\}
\end{array}
\]
These calculations are non-trivial so consider how $D_2$ is
obtained from $D_1$ by applying the clause \mbox{$\mathtt{pt(}t_1,
m, t_2, h\mathtt{) :-} 1 \separate g_4, \mathtt{=<'(}m,
x\mathtt{), pt(}xs, m, l, h\mathtt{)}$}. The following $e_i$ and
$g$ formulae are generated:
\[
\begin{array}{c@{\,}c@{\,}c@{\,}c@{\,}c}
e_3 & = & 1 \\
e_2 & = & 1 \wedge ((xs \wedge l \wedge h) \Rightarrow 1) & = & 1 \\
e_1 & = & (m \wedge x) \wedge ((m \wedge x) \Rightarrow 1) & = & m \wedge x \\
g & = & \multicolumn{3}{@{}l@{}}{1 \wedge (((t_1 \Leftrightarrow x
\wedge xs) \wedge (t_2 \Leftrightarrow x \wedge l)) \Rightarrow (m
\wedge x))}
\end{array}
\]
To characterise those \mbox{$\mathtt{pt(}t_1, m, t_2,
h\mathtt{)}$} calls which are safe, it is necessary to compute a
function $g'$ on the variables $t_1, m, t_2, h$ which, if
satisfied by the mode of a call, ensures that $g$ is satisfied by
the mode of the call. Put another way, it is necessary to
eliminate the variables $x, xs$ and $l$ from $g$ (those variables
which do not occur in the head \mbox{$\mathtt{pt(}t_1, m, t_2,
h\mathtt{)}$}) to strengthen $g$ obtain a function $g'$ such that
$g$ holds whenever $g'$ holds. This is accomplished by calculating
$g' = \forall_l \forall_{xs} \forall_{x}(g)$. First consider the
computation of $\forall_{x}(g)$:
\[
\begin{array}{r@{\;}l}
g[x \mapsto 0] = & (((t_1 \Leftrightarrow x \wedge xs) \wedge (t_2
\Leftrightarrow
x \wedge l)) \Rightarrow (m \wedge x))[x \mapsto 0] \\
 = & ((t_1 \Leftrightarrow 0 \wedge xs) \wedge (t_2
\Leftrightarrow
0 \wedge l)) \Rightarrow (m \wedge 0) \\
 = & (\neg t_1 \wedge \neg t_2) \Rightarrow 0 \\
 = & t_1 \vee t_2 \\
 \\
g[x \mapsto 1] = & (((t_1 \Leftrightarrow x \wedge xs) \wedge (t_2
\Leftrightarrow
x \wedge l)) \Rightarrow (m \wedge x))[x \mapsto 1] \\
 = & ((t_1 \Leftrightarrow xs) \wedge (t_2
\Leftrightarrow l)) \Rightarrow m \\
\end{array}
\]
Since $g[x \mapsto 0] \wedge g[x \mapsto 1] \in Pos$ it follows
that:
\[
\forall_{x}(g) = (((t_1 \Leftrightarrow xs) \wedge (t_2
\Leftrightarrow l)) \Rightarrow m) \wedge (t_1 \vee t_2)
\]
(otherwise $\forall_{x}(g)$ would be set to $0$). Eliminating the
other variables in a similar way we obtain:
\[ \begin{array}{@{}r@{\;}l@{}}
\forall_{xs} \forall_{x}(g) & = ((t_2 \Leftrightarrow l)
\Rightarrow m) \wedge
(t_1 \vee t_2) \\
g' = \forall_l \forall_{xs} \forall_{x}(g) & = m \wedge (t_1 \vee
t_2)
\end{array} \]
Observe that if $\forall_l \forall_{xs} \forall_{x}(g)$ holds then
$g$ holds. Thus if the mode of a call satisfies $g'$ then the mode
also satisfies $g$ as required. This clause thus yields the call
pattern $\langle \mathtt{pt}(\vec{w}), x_2 \wedge (x_1 \vee x_3)
\rangle$. Similarly the first and third clauses contribute the
patterns $\langle \mathtt{pt}(\vec{w}), 1 \rangle$ and $\langle
\mathtt{pt}(\vec{w}) \leftarrow x_2 \wedge (x_1 \vee x_4)
\rangle$. Observe also that
\[
1 \wedge (x_2 \wedge (x_1 \vee x_3)) \wedge (x_2 \wedge (x_1 \vee
x_4)) = x_2 \wedge (x_1 \vee (x_3 \wedge x_4))
\]
which gives the final call pattern formula for
$\mathtt{pt}(\vec{w})$ in $D_2$. The gfp is reached at $D_3$ since
$D_4 = D_3$. The gfp often expresses elaborate calling modes, for
example, it states that $\mathrm{pt}(\vec{w})$ cannot generate an
instantiation error (nor any predicate that it calls) if it is
called with its second, third and fourth argument ground. This is
a surprising result which suggests that the analysis can infer
information that might be normally missed by a programmer.

\subsubsection{Restrictions posed by the framework}\label{sect-restrict}

The chief computational requirement of the analysis is that the
input domain is equipped with a pseudo-complement operation. As
already mentioned, it is always possible to systematically design
a domain with this operator \cite{GS98} and any domain that is
known to be condensing (see section~\ref{sect-contrib}) comes
equipped with this operator. Currently, however, there are only a
few domains with a pseudo-complement. Indeed, the domain described
in \cite{CL00} appears to be unique in that it is the only type
domain that is condensing. This is the main limitation of the
backward analysis described in this paper.

$Pos$ is downward-closed in the sense that if a function $f$
describes a substitutions, then $f$ also describes all
substitutions less general than the substitution. The type domain
of \cite{CL00} is also downward-closed. It does not follow,
however, that a domain equipped with a pseudo-complement operation
is necessarily downward-closed. Heyting completion, the domain
refinement technique used to construct pseudo-complement, can be
moved to linear implication \cite{GiacobazziSR98}, though the
machinery is more complicated. However, it is likely, that in the
short term tractable condensing domains will continue to be
downward-closed. In fact, constructing tractable downward-closed
condensing domains is a topic within itself.

\section{Preliminaries}\label{sect-prelims}

\subsection{Basic Concepts}

\paragraph{Sets and sequences}
Let $\mathbb{N}$ denote the set of non-negative integers. The powerset
of $S$ is denoted $\wp(S)$. The empty sequence is denoted
$\epsilon$ and $S^{\star}$ denotes the set of (possibly empty)
sequences whose elements are drawn from $S$. Sequence
concatenation is denoted $\cdot$ and the length of a sequence $s$
is $|s|$. Furthermore, let $s^0 = \epsilon$ and \mbox{$s^{n} = s
\cdot s^{n-1}$} where $n \in \mathbb{N}$. If $n \in \mathbb{N}$
and $s \in \mathbb{N}^{\star}$ then $\max(n \cdot s) = \max(n,
\max(s))$ where $\max(\epsilon) = 0$.

\paragraph{Orderings}
A pre-order on a set $S$ is a binary relation $\sqsubseteq$ that
is reflexive and transitive. A partial order on a set $S$ is a
pre-order that is anti-symmetric. A poset $\langle S, \sqsubseteq
\rangle$ is a partial order on a set $S$. If $\langle S,
\sqsubseteq \rangle$ is a poset, then $C \subseteq S$ is a chain
iff $a \sqsubseteq b$ or $b \sqsubseteq a$ for all $a, b \in C$. A
meet semi-lattice $\langle L, \sqsubseteq, \sqcap \rangle$ is a
poset $\langle L, \sqsubseteq \rangle$ such that the meet
(greatest lower bound) $\sqcap \{ x, y \}$ exists for all $x, y
\in L$.  A complete lattice is a poset $\langle L, \sqsubseteq
\rangle$ such that the meet $\sqcap X$ and the join $\sqcup X$
(least upper bound) exist for all $X \subseteq L$. Top and bottom
are respectively defined by $\top = \sqcap \emptyset$ and $\bot =
\sqcup \emptyset$. A complete lattice is denoted $\langle L,
\sqsubseteq, \sqcap, \sqcup, \top, \bot \rangle$. Let $\langle S,
\sqsubseteq \rangle$ be a pre-order. If $X \subseteq S$ then
$\mydownarrow(X) = \{ y \in S \mid \exists x \in X . y \sqsubseteq
x \}$. If $x \in S$ then $\mydownarrow(x)$ = $\mydownarrow(\{ x
\})$. The set of order-ideals of $S$, denoted
$\wp^{\mydownarrow}(S)$, is defined by $\wp^{\mydownarrow}(S)$ =
$\{ X \subseteq S \mid X = \mydownarrow(X) \}$. Observe that
$\langle \wp^{\mydownarrow}(S), \subseteq, \cup, \cap, S,
\emptyset \rangle$ is a complete lattice.

An algebraic structure is a pair $\langle S, \mathcal{Q} \rangle$
where $S$ is a non-empty set and $\mathcal{Q}$ is collection of
$n$-ary operations $f : S^n \rightarrow S$ where $n \in
\mathbb{N}$. Let $\langle S, \sqsubseteq \rangle$ and $\langle S',
\sqsubseteq' \rangle$ be posets and $\langle S, \mathcal{Q}
\rangle$ and $\langle S', \mathcal{Q}' \rangle$ algebraic
structures such that $\mathcal{Q} = \{ f_i \mid i \in I \}$ and
$\mathcal{Q}' = \{ f'_i \mid i \in I \}$ for an index set $I$.
Then $\alpha : S \rightarrow S'$ is a semi-morphism between
$\langle S, \mathcal{Q} \rangle$ and $\langle S', \mathcal{Q}'
\rangle$ iff $\alpha(f_i(s_1, \ldots, s_n)) \sqsubseteq
f'_i(\alpha(s_1), \ldots, \alpha(s_n))$ for all $\langle s_1,
\ldots, s_n \rangle \in S^n$ and $i \in I$.

\paragraph{Functions and fixpoints}
Let $f : A \rightarrow B$. Then $\mathrm{dom}(f)$ denotes the
domain of $f$ and if $C \subseteq A$ then $f(C) = \{ f(c) \mid c
\in C \}$. Furthermore, $\mathrm{cod}(f)$ = $f(\mathrm{dom}(f))$.
Let $\langle L, \sqsubseteq, \sqcup, \sqcap \rangle$ and $\langle
L', \sqsubseteq', \sqcup', \sqcap' \rangle$ be complete lattices.
The map $f : L \rightarrow L'$ is additive iff $f(\sqcup X) =
\sqcup' f(X)$ for all $X \subseteq L$; $f$ is continuous iff
$f(\sqcup C) = \sqcup' f(C)$ for all chains $C \subseteq L$; $f$
is co-continuous iff $f(\sqcap C) = \sqcap' f(C)$ for all chains
$C \subseteq L$ and $f$ is monotonic iff $f(x) \sqsubseteq' f(y)$
for all $x \sqsubseteq y$. Let $x \sqsubseteq y$. If $f$ is
continuous then $f(y) = f(x \sqcup y) = \sqcup' \{ f(x), f(y) \}$
and thus $f(x) \sqsubseteq' f(y)$. If $f$ is co-continuous then
$f(x) = f(x \sqcap y) = \sqcap' \{ f(x), f(y) \}$ and thus $f(x)
\sqsubseteq' f(y)$. Both continuity and co-continuity thus imply
monotonicity. If $f : L \rightarrow L$, then $f$ is idempotent iff
$f(x) = f^{2}(x)$ for all $x \in L$ and $f$ is extensive iff $x
\sqsubseteq f(x)$ for all $x \in L$. The Knaster-Tarski theorem
states that any monotone operator $f : L \rightarrow L$ on a
complete lattice $\langle L, \sqsubseteq, \sqcup, \sqcap, \top,
\bot \rangle$ admit both greatest and least fixpoints that are
characterised by $\mathrm{gfp}(f) = \sqcup \{ x \in L \mid x
\sqsubseteq f(x) \}$ and $\mathrm{lfp}(f) = \sqcap \{ x \in L \mid
f(x) \sqsubseteq x \}$. If $f$ is co-continuous then
$\mathrm{gfp}(f) = \sqcap_{n \in \nat} f^{n}(\top)$ and dually if
$f$ is continuous then $\mathrm{lfp}(f) = \sqcup_{n \in \nat}
f^{n}(\bot)$. $\{ f^{n}(\top) \mid n \in \nat \}$ and $\{
f^{n}(\bot) \mid n \in \nat \}$ are, respectively, the lower and
upper Kleene iteration sequences of $f$.

\paragraph{Galois insertions and closure operators}
If $\langle S, \sqsubseteq \rangle$ and $\langle S', \sqsubseteq'
\rangle$ are posets and \linebreak \mbox{$\alpha : S \rightarrow
S'$} and $\gamma : S' \rightarrow S$ are monotonic maps such that
$\forall x \in S . x \sqsubseteq \gamma(\alpha(x))$ and $\forall
x' \in S' . \alpha(\gamma(x')) \sqsubseteq' x'$, then the
quadruple $\langle S, \gamma, S', \alpha \rangle$ is a Galois
connection between $S$ and $S'$. In other words, $\alpha$ is the
lower (or left) adjoint of $\gamma$ and $\gamma$ is the upper (or
right) adjoint of $\alpha$. If, in addition, $\forall x' \in S' .
x' \sqsubseteq' \alpha(\gamma(x'))$, then $\langle S, \gamma, S',
\alpha \rangle$ is a Galois insertion between $S$ and $S'$. The
operator $\rho : L \rightarrow L$ on a complete lattice $\langle
L, \sqsubseteq \rangle$ is a closure operator iff $\rho$ is
monotonic, idempotent and extensive. The set of closure operators
on $L$ is denoted $uco(L)$. The image set $\rho(L)$ of a closure
operator $\rho$ is a complete lattice with respect to
$\sqsubseteq$. A Galois insertion $\langle L, \gamma, L', \alpha
\rangle$ between the complete lattices $L$ and $L'$ defines the
closure operator $\rho = \gamma \circ \alpha$. Conversely, a
closure operator $\rho : L \rightarrow L$ on the complete lattice
$\langle L, \sqsubseteq, \sqcup \rangle$ defines the Galois
insertion $\langle L, id, \rho(L), \rho \rangle$ where $id$
denotes identity. Galois insertions and closure operators are thus
isomorphic, though closure operators are typically more succinct
and hence used in this paper.

\paragraph{Substitutions}
Let $Sub$ denote the set of (idempotent) substitutions and
let $Ren$
denote the set of (bijective) renaming substitutions.

\subsection{Cylindric constraint systems}

Let $V$ denote a (denumerable) universe of variables and let
$\mathcal{C}$ denote a constraint system over $V$. An algebra
$\langle \mathcal{C}, \mymodels, \mywedge, 1, \{ \exists_x \}_{ x
\in V }, \{ d_{x, y} \}_{ x, y \in V } \rangle$ is a
semi-cylindric constraint system iff $\langle \mathcal{C},
\mymodels, \mywedge \rangle$ is a meet semi-lattice with a top
$1$; $\exists_x$ is a family of (unary) cylindrification
operations such that: $c \mymodels \exists_x(c)$, $\exists_x(c)
\mymodels \exists_x(c')$ if $c \mymodels c'$, $\exists_x(c
\mywedge \exists_x (c')) = \exists_x(c) \mywedge \exists_x(c')$;
and $d_{x,y}$ is a family of (constant) diagonalisation operations
such that: $d_{x,x} = 1$, $d_{x,y} = \exists_z(d_{x,z} \mywedge
d_{z, y})$ and $d_{x, y} \mywedge \exists_x(c \mywedge d_{x, y})
\mymodels c$ if $x \neq y$. Cylindrification captures the concept
of projecting out a variable (and is useful in modeling variables
that go out of scope) whereas diagonalisation captures the notion
of an alias between two variables (and is useful in modeling
parameter passing). (The reader is referred to \cite{GDL95} for
further details on cylindric constraint systems and their
application in abstract interpretation.)

\begin{example}\label{exam-herbrand} \rm
An equation $e$ is a pair $(s = t)$ where $s$ and $t$ are terms. A
finite conjunction of equations is denoted $E$ and $Eqn$ denotes
the set of finite conjunctions of equations. Let $eqn(\theta) = \{
x = t \mid x \mapsto t \in \theta \}$ and $unify(E) = \{ \theta
\in Sub \mid \forall (s = t) \in E . \theta(s) = \theta(t) \}$.
$Eqn$ is pre-ordered by entailment $E_1 \mymodels E_2$ iff
$unify(E_1) \subseteq unify(E_2)$ and quotiented by $E_1 \approx
E_2$ iff $E_1 \mymodels E_2$ and $E_2 \mymodels E_1$. This gives
the meet semi-lattice $\langle Eqn/\!\!\approx, \mymodels,
\mywedge \rangle$ with a top $1$ where conjunction is defined
$[E_1]_{\approx} \mywedge [E_2]_{\approx} = [E_1 \cup
E_2]_{\approx}$ and $1 = [\emptyset]_{\approx}$. Let $mgu(E)$ =
$\{ \theta \in unify(E) \mid \forall \kappa \in unify(E) \, . \,
eqn(\kappa) \mymodels eqn(\theta) \}$. Finally, let $d_{x,y} = [\{
x = y \}]_{\approx}$ and define project out by
$\exists_x([E]_{\approx}) = [eqn(\{ y \mapsto t \in \theta \mid x
\neq y \})]_{\approx}$ if $\theta \in mgu(E)$. Otherwise,
if $mgu(E) = \emptyset$,
define $\exists_x([E]_{\approx}) = [\{ a = b \}]_{\approx}$ where $a$ and
$b$ are distinct constant symbols. Then \linebreak $\langle Eqn/\!\!\approx,
\mymodels, \mywedge, 1, \{ \exists_x \}_{ x \in V }, \{ d_{x, y}
\}_{ x, y \in V } \rangle$ is a semi-cylindric constraint system.
\end{example}

\noindent An algebra $\langle \mathcal{C}, \mymodels, \myvee, \mywedge,
1, 0, \{ \exists_x \}_{ x \in V }, \{ d_{x, y} \}_{ x, y \in V }
\rangle$ that extends a semi-cylindric constraint system to a
complete lattice $\langle \mathcal{C}, \mymodels, \myvee, \mywedge, 1, 0
\rangle$ is a cylindric constraint system. A semi-cylindric
constraint system can be lifted to a cylindric constraint system
via a power-domain construction. In particular $\langle
\wp^{\mydownarrow}(\mathcal{C}), \subseteq, \cup, \cap,
\mathcal{C}, \emptyset$, $\{ \exists'_x \}_{ x \in V }$, $\{ d'_{x, y}
\}_{ x, y \in V } \rangle$ is a cylindric constraint system where
$\exists'_x(C) = \mydownarrow(\{ \exists_x(c) \mid c \in C \})$
and $d'_{x, y} = \mydownarrow(d_{x, y})$.

\begin{example} The semi-cylindric system
of example~\ref{exam-herbrand} can be lifted to the cylindric
system $\langle \wp^{\downarrow}(Eqn), \subseteq, \cup, \cap, Eqn,
\emptyset, \exists', d' \rangle$ where $\exists'_x(C) \! = \!
\mydownarrow(\{ \exists_x(c) | c \in C \})$ and $d'_{x, y} \! = \!
\mydownarrow(d_{x, y})$.
\end{example}

In the sequel, unless otherwise stated, all constraint systems
considered are over the same $V$ and thus a cylindric constraint
system will be simply denoted \linebreak $\langle \mathcal{C},
\mymodels, \myvee, \mywedge, 1, 0, \exists, d \rangle$. Let
$\mathrm{var}(o)$ denote the set of the variables in the syntactic
object $o$ and let $FV(c)$ denote the set of free variables in a
constraint $c \in \mathcal{C}$, that is, $FV(c) = \{ x \in
\mathrm{var}(c) \mid \exists y \in V \, . \, c \neq \exists_{x}(c
\mywedge d_{x, y}) \}$. Abbreviate project out by $\exists_{\{
x_1, \ldots, x_n \} }(c)$ = $\exists_{x_1}( \ldots
(\exists_{x_n}(c)))$ and project onto by
$\overline{\exists}_{X}(c)$ = $\exists_{FV(c) \setminus X}(c)$.
Let $d_{\vec{x}, \vec{y}} = \mywedge_{i = 1}^{n} d_{x_i, y_i}$
where $\vec{x} = \langle x_1 \ldots x_n \rangle$ and $\vec{y} =
\langle y_1 \ldots y_n \rangle$. If $c \in \mathcal{C}$ then let
$\partial_{\vec{x}}^{\vec{y}}(c)$ denote the constraint obtained
by replacing $\vec{x}$ with $\vec{y}$, that is,
$\partial_{\vec{x}}^{\vec{y}}(c)$ =
\mbox{$\exists_{\vec{z}}(\exists_{\vec{x}}(c \mywedge d_{\vec{x},
\vec{z}}) \mywedge d_{\vec{z}, \vec{y}})$} where
$\mathrm{var}(\vec{z}) \cap (FV(c) \cup \mathrm{var}(\vec{x}) \cup
\mathrm{var}(\vec{y})) = \emptyset$. Finally, if $C \subseteq
\mathcal{C}$ then $\partial_{\vec{x}}^{\vec{y}}(C)$ = \mbox{$\{
\partial_{\vec{x}}^{\vec{y}}(c) \mid c \in C \}$}.

\begin{example}\label{exam-EPos} \rm
Let $X$ be a finite subset of $V$. The groundness domain $\langle
EPos_X, \models, \curlyvee, \wedge, 1, 0 \rangle$ \cite{HACK00} is
a finite lattice where $EPos_X = \{ 0 \} \cup \{ \wedge F \mid F
\subseteq X \cup E_X \}$, $E_X = \{ x \Leftrightarrow y \mid x, y
\in X \}$ and $f_1 \curlyvee f_2 = \wedge \{ f \in EPos_X \mid f_1
\models f \wedge f_2 \models f \}$. $EPos_X$ is a cylindric
constraint system with $d_{x, y} = (x \Leftrightarrow y)$ and
$\exists_x(f) = f' \wedge f''$ where $f' = \wedge \{ y \in Y \mid
f \models y \}$, $f'' = \wedge \{ e \in E_{Y} \mid f \models e \}$
and $Y = X \setminus \{ x \}$.
\end{example}

\begin{example}\label{exam-Pos} \rm Let $Bool_X$ denote the
Boolean functions over $X$. The dependency domain $Pos_X$
\cite{AMSS98} is defined by $Pos_X = \{ 0 \} \cup \{ f \in Bool_X
\mid \wedge X \models f \}$. Henceforth $Y$ abbreviates $\wedge
Y$. The lattice $\langle Pos_X, \models, \vee, \wedge, 1, 0
\rangle$ is finite and is a cylindric constraint system with
$d_{x, y} = (x \Leftrightarrow y)$ and Schr\"oder elimination
defining $\exists_x(f) = f[x \mapsto 1] \vee f[x \mapsto 0]$.
\end{example}

\subsection{Complete Heyting algebras}

Let $\langle L, \sqsubseteq, \sqcap \rangle$ be a lattice with $x,
y \in L$. The pseudo-complement of $x$ relatively to $y$, if it
exists, is a unique element $z \in L$ such that $x \sqcap w
\sqsubseteq y$ iff $w \sqsubseteq z$. $L$ is relatively
pseudo-completed iff the pseudo-complement of $x$ relative to $y$,
denoted $x \rightarrow y$, exists for all $x, y \in L$. If $L$ is
also complete then it is a complete Heyting algebra (cHa). If $x,
y \in L$ then $x \sqcap (x \rightarrow y) = x \sqcap y$.
Furthermore, if $\langle L, \sqsubseteq, \sqcup, \sqcap \rangle$
is a cHa then $x \rightarrow y = \sqcup \{ w \in L \mid x \sqcap w
\sqsubseteq y \}$. The intuition behind the pseudo-complement of
$x$ relative to $y$ is that it is the weakest element whose
combination (meet) with $x$ implies $y$. Interestingly
pseudo-complement can be interpreted as the adjoint of
conjunction. (The reader is referred to \cite{VanDalen} for
further details on complete Heyting algebras.) The following
result \cite{B67}[Chapter~IX, Theorem~15] explains how a cHa
depends on the additivity of meet.

\begin{theorem}\label{theorem-dist} \rm A complete lattice $L$
is relatively pseudo-complemented iff
$x \sqcap (\sqcup Y) = \sqcup \{ x \sqcap y \mid y \in Y \}$ for all
$x \in L$ and $Y \subseteq L$.
\end{theorem}

\begin{example}\label{exam-pseudo} \rm
Let $\{ x, y \} \subseteq X$ and $f = (x \Leftrightarrow y)$. Then
returning to $EPos_X$ of example~\ref{exam-EPos},
$f \wedge (\curlyvee \{ x, y \}) = f \wedge (1) = f \neq (x
\wedge y) = \curlyvee \{ x \wedge y, x \wedge y \} = \curlyvee \{
f \wedge x, f \wedge y \}$. Hence, by theorem~\ref{theorem-dist},
$EPos_X$ is not a cHa. Now consider $Pos_X$ of example~\ref{exam-Pos},
and specifically let $f \in Pos_X$ and $G \subseteq
Pos_X$. Since $\wedge$ distributes over $\vee$, it follows that $f
\sqcap (\sqcup G) = \sqcup \{ f \sqcap g \mid g \in G \}$, thus by
theorem~\ref{theorem-dist}, $Pos_X$ is a cHa. Similarly, $\cap$
distributes over $\cup$, and thus it follows by
theorem~\ref{theorem-dist} that $\wp^{\downarrow}(\mathcal{C})$ is
also a cHa.
\end{example}

%

\subsection{Constraint logic programs}

Let $\mathrm{\Pi}$ denote a (finite) set of predicate symbols, let
$Atom$ denote the set of (flat) atoms over $\mathrm{\Pi}$ with
distinct arguments drawn from $V$, and let $\langle \mathcal{C},
\mymodels, \myvee, \mywedge, 1, 0, \exists, d \rangle$ be a
semi-cylindric constraint system. The set of constrained atoms is
defined by $Base^{\mathcal{C}}$ = $\{ p(\vec{x}) \neck c \mid
p(\vec{x}) \in Atom \wedge c \in \mathcal{C} \}$. Let
$FV(p(\vec{x}) \neck c) = \mathrm{var}(\vec{x}) \cup FV(c)$.
Entailment $\mymodels$ lifts to $Base^{\mathcal{C}}$ by $w_1
\mymodels w_2$ iff $\overline{\exists}_{\vec{x}}(d_{\vec{x},
\vec{x}_{1}} \mywedge c_1) \mymodels
\overline{\exists}_{\vec{x}}(d_{\vec{x}, \vec{x}_{2}} \mywedge
c_2)$ where \mbox{$w_i = p(\vec{x}_{i}) \neck c_i$} and
$\mathrm{var}(\vec{x}) \cap (FV(w_1) \cup FV(w_2)) = \emptyset$.
This pre-order defines the equivalence relation $w_1 \approx w_2$
iff $w_1 \mymodels w_2$ and $w_2 \mymodels w_1$ to give a set of
interpretations defined by $Int^{\mathcal{C}}$ =
$\wp(Base^{\mathcal{C}} / \!\!\approx)$. $Int^{\mathcal{C}}$ is
ordered by $I_1 \sqsubseteq I_2$ iff for all $[w_1]_{\approx} \in
I_1$ there exists $[w_2]_{\approx} \in I_2$ such that $w_1
\mymodels w_2$. Let $\equiv$ denote the induced equivalence
relation $I_1 \equiv I_2$ iff $I_1 \sqsubseteq I_2$ and $I_2
\sqsubseteq I_1$. $\langle Int^{\mathcal{C}}/\!\!\equiv,
\sqsubseteq, \sqcup, \sqcap, \top, \bot \rangle$ is a complete
lattice where $[I_1]_{\equiv} \sqcup [I_2]_{\equiv}$ = $[I_1 \cup
I_2]_{\equiv}$, $[I_1]_{\equiv} \sqcap [I_2]_{\equiv}$ = $[\cup \{
I \mid I \sqsubseteq I_1 \wedge I \sqsubseteq I_2 \}]_{\approx}$,
$\top = [\{ [{p(\vec{x}) \neck 1}]_{\approx} \mid p(\vec{x}) \in
Atom \}]_{\equiv}$ and $\bot = [\emptyset]_{\equiv}$.

A constraint logic program $P$ over $\mathcal{C}$ is a finite set
of clauses $w$ of the form $w = h \neck c, g$ where $h \in Atom$,
$c \in \mathcal{C}$, $g \in Goal$ and $Goal = Atom^{\star}$. The
fixpoint semantics of $P$ is defined in terms of an immediate
consequences operator $\mathcal{F}_P^\mathcal{C}$.

\begin{definition} \rm
Given a constraint logic program $P$ over a semi-cylindric
constraint system $\mathcal{C}$, the operator
$\mathcal{F}_P^\mathcal{C} : Int^{\mathcal{C}} \rightarrow
Int^{\mathcal{C}}$ is defined by:
\[
\mathcal{F}_P^\mathcal{C}(I) = \left\{ [p(\vec{x}) \neck
c']_{\approx} \left|
\begin{array}{ccc}
\exists & p(\vec{x}) \neck c, p_{1}(\vec{x}_1), \ldots, p_{n}(\vec{x}_n) \in P & . \\
\exists & \{ [{p_{i}(\vec{x}_i) \neck c_i}]_{\approx} \}_{i=1}^{n} \subseteq I & . \\
        & c' = c \mywedge \mywedge_{i = 1}^{n} {\overline{\exists}_{\vec{x_i}}}(c_i) &
\end{array}
\right. \right\}
\]
\end{definition}

\noindent The operator $\mathcal{F}_P^\mathcal{C}$ lifts to
$Int^{\mathcal{C}}/\!\!\equiv$ by
$\mathcal{F}_P^\mathcal{C}([I]_{\equiv}) =
[\mathcal{F}_P^\mathcal{C}(I)]_{\equiv}$. The lifting is monotonic
and hence the fixpoint semantics for a program $P$ over
$\mathcal{C}$ exists and is denoted $\mathcal{F}^\mathcal{C}(P) =
\mathrm{lfp}(\mathcal{F}_P^\mathcal{C})$. (The reader is referred
to \cite{BGLM94,JM94} for further details on semantics and
constraint logic programming.)

The operational semantics of $P$ is defined in terms of a
transition system $\reduce_{P}$ between states of the form $State
= Goal \times \mathcal{C}$. To define the transition system, let
$FV(\langle g; c \rangle)$ = $\mathrm{var}(g) \cup FV(c)$ and
$FV(h \neck c, g)$ = $\mathrm{var}(h) \cup FV(c) \cup
\mathrm{var}(g)$. To rename clauses with $\varphi \in Ren$ it is
necessary to rename constraints with $\varphi$. Thus define
$\varphi(h \neck c, g)$ = $\varphi(h) \neck
\partial_{\vec{x}}^{\varphi(\vec{x})}(c), \varphi(g)$. To rename
apart from a syntactic object $o$, let $w \ll_{o} P$ indicate that
there exists $w' \in P$ and $\varphi \in Ren$ such that
$\mathrm{var}(\mathrm{cod}(\varphi)) \cap FV(w') = \emptyset$,
$\varphi(w') = w$ and $FV(o) \cap FV(w) = \emptyset$.

\begin{definition} \rm
Given a constraint logic program $P$ over a semi-cylindric
constraint system $\mathcal{C}$, \linebreak $\reduce_{P} \subseteq
State^2$ is the least relation such that:
\[ s = \langle p(\vec{x}), g; c \rangle
\reduce_P \langle g', g; c \mywedge d_{\vec{x}, \vec{x}'} \mywedge
c' \rangle \] where $p(\vec{x}') \neck c', g' \ll_{s} P$.
\end{definition}

\noindent The operational semantics is specified by the transitive
closure of the transition relation on (atomic) goals, that is,
$\mathcal{O}^\mathcal{C}(P) = [\{ [{p(\vec{x}) \neck c}]_{\approx}
\mid \langle p(\vec{x}); 1 \rangle \reduce^{\star}_{P} \langle
\epsilon; c \rangle \}]_{\equiv}$. The relationship between the
operational and fixpoint semantics is stated below.

\begin{theorem}\label{theorem-equiv} \rm
$\mathcal{O}^\mathcal{C}(P) = \mathcal{F}^\mathcal{C}(P)$.
\end{theorem}

\subsection{Abstract semantics for constraint logic programs}

To apply abstraction techniques and finitely characterise
$\mathcal{F}^\mathcal{C}(P)$, and thereby
$\mathcal{O}^\mathcal{C}(P)$, the semi-cylindric domain
$\mathcal{C}$ is replaced by the cHa
$\wp^{\mydownarrow}(\mathcal{C})$ which is particularly amenable
to approximation and backward reasoning.

If $P$ is a constraint logic program over $\mathcal{C}$, then
$\mydownarrow(P) = \{ h \neck \mydownarrow(c), g \mid h \neck c, g
\in P \}$. Furthermore, if $I \in Int^{\mathcal{C}}$, then let
$\mydownarrow({[I]_{\equiv}}) = [{\{ [{p(\vec{x}) \neck \;
\mydownarrow(c)}]_{\approx} \mid [{p(\vec{x}) \neck c}]_{\approx}
\in I \}}]_{\equiv}$. Note the overloading on $\approx$ and hence
$\equiv$. The $\approx$ of $[{p(\vec{x}) \neck c}]_{\approx}$ is
induced by $\langle \mathcal{C}, \mymodels \rangle$ whereas the
$\approx$ of $[{p(\vec{x}) \neck \; \mydownarrow(c)}]_{\approx}$
is induced by $\langle \wp^{\mydownarrow}(\mathcal{C}), \subseteq
\rangle$. The following proposition details the relationship
between $\mathcal{F}^\mathcal{C}$ and
$\mathcal{F}^{\wp^{\mydownarrow}(\mathcal{C})}$.

\begin{proposition}\label{prop-down} \rm
$\mydownarrow(\mathcal{F}^\mathcal{C}(P)) \sqsubseteq
\mathcal{F}^{\wp^{\mydownarrow}(\mathcal{C})}(\mydownarrow(P))$.
\end{proposition}

Let $\langle \mathcal{C}, \mymodels, \myvee, \mywedge, 1, 0,
\exists, d \rangle$ denote a cylindric constraint system. If $\rho
\in uco(\mathcal{C})$ then $\langle \rho(\mathcal{C}), \mymodels,
\mywedge \rangle$ is a complete lattice. If $\rho$ is additive,
then $\langle  \rho(\mathcal{C}), \mymodels, \myvee, \mywedge
\rangle$ is a sub-lattice of $\langle \mathcal{C}, \mymodels,
\myvee, \mywedge \rangle$. More generally, the join is denoted
$\myvee'$. Observe that $\rho(\mathcal{C})$ has $1$ and $\rho(0)$
for top and bottom and $c_1 \myvee c_2 \mymodels \rho(c_1 \myvee
c_2) = c_1 \myvee' c_2$ for all $c_1, c_2 \in \rho(\mathcal{C})$.
A cylindric constraint system is obtained by augmenting
$\rho(\mathcal{C})$ with cylindrification $\exists'_x$ and
diagonalisation $d'_{x, y}$ operators. To abstract $\langle
\mathcal{C}, \mymodels, \myvee, \mywedge, 1, 0, \exists, d
\rangle$ safely with $\langle \rho(\mathcal{C}), \mymodels,
\myvee', \mywedge, 1, \rho(0), \exists', d' \rangle$, $\rho$ is
required to be a semi-morphism \cite{GDL95} which additionally
requires that $\rho(\exists_x(c)) \mymodels {\exists'_x}(\rho(c))$
for all $c \in \mathcal{C}$ and \linebreak \mbox{$\rho(d_{x, y})
\mymodels d'_{x, y}$} for all \mbox{$x, y \in V$}. In fact, these
requirements turn out to be relatively weak conditions: most
abstract domains come equipped with (abstract) operators to model
projection and parameter passing.

\begin{example}\label{exam-ground} \rm
Consider the cylindric system $\langle \wp^{\downarrow}(Eqn),
\subseteq, \cup, \cap, Eqn, \emptyset, \exists, d \rangle$ derived
from the semi-cylindric system introduced in
example~\ref{exam-herbrand}. Let $Bool = Bool_V$ and $Pos =
Pos_V$. Define \mbox{$\alpha_{Pos} : \wp^{\downarrow}(Eqn)
\rightarrow Pos$} by $\alpha_{Pos}(C) = \vee \{ \alpha(\theta)
\mid \theta \in mgu(E) \wedge E \in C \}$ and $\alpha(\theta)$ =
\mbox{$\wedge \{ x \Leftrightarrow \mathrm{var}(t) \! \mid \! x
\mapsto t \in \theta \}$}. Also define $\gamma_{Pos} : Pos
\rightarrow \wp^{\downarrow}(Eqn)$ by \mbox{$\gamma_{Pos}(f) =
\cup \{ C \in \wp^{\downarrow}(Eqn) \mid \alpha_{Pos}(C) \models f
\}$} and observe $\rho_{Pos} \in uco(\wp^{\downarrow}(Eqn))$ where
$\rho_{Pos} = \gamma_{Pos} \circ \alpha_{Pos}$. To construct a
semi-morphism, put $d'_{x, y} = \gamma_{Pos}(x \Leftrightarrow y)$
and $\exists'_x(C) = \gamma_{Pos}(f[x \mapsto 1] \vee f[x \mapsto
0])$ where $f = \alpha_{Pos}(C)$. Then $\rho_{Pos}({d_{x, y}})
\subseteq d'_{x, y}$ and $\rho_{Pos}(\exists_x(C)) \subseteq
\exists'_x(\rho_{Pos}(C))$ for all $C \in
\wp^{\mydownarrow}(Eqn)$. Note that $C_1 \cap C_2 =
\gamma_{Pos}(f_1 \wedge f_2)$ and $C_1 \myvee' C_2 =
\gamma_{Pos}(f_1 \vee f_2)$ where $C_i = \gamma_{Pos}(f_i)$.
Surprisingly $C_1 \myvee' C_2 \neq C_1 \cup C_2$ \cite{FR94}, as
witnessed by $C_1 = \gamma_{Pos}(x)$ and $C_2 = \gamma_{Pos}(x
\Leftrightarrow y)$ since $\{y = f(x,z)\} \not\in C_1 \cup C_2$
whereas $\alpha_{Pos}(\{ y = f(x,z) \}) = y \Leftrightarrow (x
\wedge z) \models x \vee (x \Leftrightarrow y)$ so that $\{y =
f(x,z)\} \in C_1 \myvee' C_2 = \gamma_{Pos}(x \vee (x
\Leftrightarrow y))$. Nevertheless, $\rho_{Pos}$ is a
semi-morphism between $\langle \wp^{\downarrow}(Eqn), \subseteq,
\cup, \cap, Eqn, \emptyset, \exists, d \rangle$ and $\langle
\rho_{Pos}(\wp^{\mydownarrow}(Eqn)), \subseteq, \myvee', \cap,
Eqn, \rho_{Pos}(\emptyset), \exists', d' \rangle$.
\end{example}

\noindent The operator $\rho$ lifts to the complete lattice
$Int^{\mathcal{C}}/\!\!\equiv$ by $\rho([I]_{\equiv}) =
[\rho(I)]_{\equiv}$ where $\rho(I) = \{ [{p(\vec{x}) \neck
\rho(c)}]_{\approx} \mid [{p(\vec{x}) \neck c}]_{\approx} \in I
\}$. Thus $\rho \in uco(Int^{\mathcal{C}}/\!\!\equiv)$. It is also
useful to lift $\rho$ to programs by $\rho(P) = \{ h \neck
\rho(c), g \mid h \neck c, g \in P \}$. The following result
relates the fixpoint semantics of $P$ to that of its abstraction
$\rho(P)$.

\begin{theorem}\label{theorem-approx} \rm
Let ${\mathcal{C}}$ be a cylindric constraint system. If $\rho \in
uco(\mathcal{C})$ is a semi-morphism, then
$\rho(\mathcal{F}^{\mathcal{C}}(P)) \sqsubseteq
\mathcal{F}^\mathcal{\mathrm{cod}(\rho)}(\rho(P))$.
\end{theorem}

\begin{corollary}\label{cor-semi} \rm
Let ${\mathcal{C}}$ be a semi-cylindric constraint system. If
$\rho \in uco(\wp^{\mydownarrow}(\mathcal{C}))$ is a
semi-morphism, then
$\rho(\mydownarrow(\mathcal{F}^{\mathcal{C}}(P))) \sqsubseteq
\mathcal{F}^{\mathrm{cod}(\rho)}(\rho(\mydownarrow(P)))$.
\end{corollary}

\section{Constraint logic programs with assertions}\label{sect-assert}

We consider programs annotated with assertions \cite{DM88}. When
considering the operational semantics of a constraint logic
program, it is natural to associate assertions with syntactic
elements of the program such as predicates or the program points
between body atoms. Without loss of generality, we decorate the
neck of each clause with a set of constraints $C$ that is
interpreted as an assertion. When $C$ is encountered, the store
$c$ is examined to determine whether $c \in C$ (modulo renaming).
If $c \in C$ execution proceeds normally, otherwise an error
state, denoted $\lozenge$, is entered and execution halts.

To formalise this idea, let $\mathcal{C}$ be a semi-cylindric
constraint system and \linebreak \mbox{$\rho \in
uco(\wp^{\mydownarrow}(\mathcal{C}))$}. The assertion language (in
whatever syntactic form it takes) is described by $\rho$. A clause
of a constraint logic program over $\mathcal{C}$ with assertions
over $\mathrm{cod}(\rho)$ then takes the form $h \neck C \separate
c, g$ where $h \in Atom$, $C \in \mathrm{cod}(\rho)$, $c \in
\mathcal{C}$, $g \in Goal$ and $\separate$ separates the assertion
from the body of the clause. Notice that $C$ is an order-ideal and
thus downward closed. ($C$ can thus represent disjunctions of
constraints, but the semantics presented in this section should
not be confused with a collecting semantics.) Note also that
program transformation \cite{PBH00a} can be used to express
program point assertions in terms of our assertion language. To
specify the behaviour of programs with assertions, let
$State_{\lozenge} = State \cup \{ \lozenge \}$, and let
$\mathrm{CLP}(P) = \{ h \neck c, g \mid h \neck C \separate c, g
\in P \}$. The following definition details how the operational
semantics for the assertion language is realised in terms of
projection, renaming and a test for inclusion.

\begin{definition} \rm Given a constraint logic program $P$ over a
semi-cylindric constraint system $\mathcal{C}$ with assertions
over $\rho(\wp^{\mydownarrow}(\mathcal{C}))$, \mbox{$\rreduce_{P}
\subseteq State \times State_{\lozenge}$} is the least relation
such that:
\[
s = \langle p(\vec{x}), g; c \rangle \rreduce_{P}
\left\{ \begin{array}{ll@{\;}l}

\lozenge & \mathrm{if} & p(\vec{x}') \neck C' \separate c', g' \in
P
\\
& \wedge & \partial_{\vec{x}}^{\vec{x}'}(\overline{\exists}_{\vec{x}}(c)) \not\in C' \\

\langle g', g; c \mywedge d_{\vec{x}, \vec{x}'} \mywedge c'
\rangle & \mathrm{else} & \mathrm{if} \; p(\vec{x}') \neck c', g'
\ll_{s} \mathrm{CLP}(P)

\end{array} \right.
\]
\end{definition}

\noindent Recall that $p(\vec{x}') \neck c', g' \ll_{s}
\mathrm{CLP}(P)$ ensures that the clause $p(\vec{x}') \neck c',
g'$ does not share any variables with $s$. The operational
semantics of $P$ is then defined in terms of
$\rreduce_{P}^{\star}$ as $\mathcal{A}^{\rho, \mathcal{C}}(P) =
[\{ [{p(\vec{x}) \neck c}]_{\approx} \mid \langle p(\vec{x}); 1
\rangle \rreduce^{\star}_{P} \langle \epsilon; c \rangle
\}]_{\equiv}$. The relationship between two operational semantics
is stated in the following (trivial) result.

\begin{proposition}\label{prop-trivial} \rm
$\mathcal{A}^{\rho, \mathcal{C}}(P) \sqsubseteq
\mathcal{O}^\mathcal{C}(\mathrm{CLP}(P))$
\end{proposition}

\noindent Assertions are often used as interface between behaviour
that is amenable to formalisation, for example as an operational
semantics, and behaviour that is less tractable, for example, the
semantics of a builtin \cite{PBH00b}. More to the point, it is not
always possible to infer the behaviour of a builtin from its
definition, partly because builtins are often complicated and
partly because builtins are often expressed in a language such as
C. Our work requires assertions for each builtin in order to
specify: its calling convention (for example, which arguments are
required to be ground) and its success behaviour (for example,
which arguments are grounded).

\section{Backward fixpoint semantics for constraint logic programs
with assertions}\label{sect-collect}

Let $P$ be a constraint logic program over the semi-cylindric
constraint system $\mathcal{C}$ with assertions over
$\rho(\wp^{\mydownarrow}(\mathcal{C}))$. One natural and
interesting question is whether the error state $\lozenge$ is
reachable (or conversely not reachable) in $P$ from an initial
state $\langle p(\vec{x}); c \rangle$. For a given constraint
logic program $P$ with assertions, the backward fixpoint semantics
presented in this section infers a (possibly empty) set of $c \in
\mathcal{C}$ for which $\langle p(\vec{x}); c \rangle
\not\rreduce^{\star}_{P} \lozenge$. The semantics formalises the
informal backward analysis sketched in
section~\ref{sect-informal}.

For generality, the semantics is parameterised by $\mathcal{C}$
and $\rho$. The correctness argument requires $\rho$ to be a
semi-morphism between \mbox{$\langle
\wp^{\downarrow}(\mathcal{C}), \subseteq, \cup, \cap, \mathcal{C},
\emptyset, \exists, d \rangle$} and \mbox{$\langle
\rho(\wp^{\mydownarrow}(\mathcal{C})), \subseteq, \myvee', \cap,
\mathcal{C}, \rho(\emptyset), \exists', d' \rangle$}.
Additionally, $\rho(\wp^{\mydownarrow}(\mathcal{C}))$ must be a
cHa, that is, it must possess a pseudo-complement $\rightarrow'$.
To explain, how pseudo-complement aids backward analysis consider
the problem of inferring $c \in \mathcal{C}$ for which $\langle g;
c \rangle \not\rreduce^{\star}_{P} \lozenge$ where $g =
p_1(\vec{x}_1), \ldots, p_n(\vec{x}_n)$. Suppose $f_i \in
\rho(\wp^{\mydownarrow}(\mathcal{C}))$ describes the success
pattern for $p_i(\vec{x}_i)$, that is, if $\langle p_i(\vec{x}_i);
1 \rangle \reduce^{\star}_{P} \langle \epsilon; c \rangle$ then $c
\in f_i$. Moreover, suppose $d_i \in
\rho(\wp^{\mydownarrow}(\mathcal{C}))$ approximates the initial
call pattern for $p_i(\vec{x}_i)$, that is, if $c \in d_i$ then
$\langle p_i(\vec{x}_i); c \rangle \not\rreduce^{\star}_{P}
\lozenge$. Observe that $\langle p_{n-1}(\vec{x}_{n-1}),
p_n(\vec{x}_n); c \rangle \not\rreduce^{\star}_{P} \lozenge$ if $c
\in d_{n-1} \cap e$ and $e \cap (d_{n-1} \cap f_{n-1}) \subseteq
d_n$. This follows since $\langle p_{n-1}(\vec{x}_{n-1}); c
\rangle \not\rreduce^{\star}_{P} \lozenge$ because $c \in d_{n-1}
\cap e \subseteq d_{n-1}$. Moreover, if $\langle
p_{n-1}(\vec{x}_{n-1}), p_n(\vec{x}_n); c \rangle
\rreduce^{\star}_{P} \langle p_n(\vec{x}_n); c' \rangle$ then $c'
\in (d_{n-1} \cap e) \cap f_{n-1} \subseteq d_n$ and thus $\langle
p_{n}(\vec{x}_{n}); c' \rangle \not\rreduce^{\star}_{P} \lozenge$.
Putting $e = \rho(\emptyset)$ ensures $e \cap (d_{n-1} \cap
f_{n-1}) \subseteq d_n$ and thereby achieves correctness. However,
for precision, $d_{n-1} \cap e$ should be maximised. Since
$\rho(\wp^{\mydownarrow}(\mathcal{C}))$ is a cHa, this reduces to
assigning $e = \myvee' \{ e' \in
\rho(\wp^{\mydownarrow}(\mathcal{C})) \mid e' \cap (d_{n-1} \cap
f_{n-1}) \subseteq d_n \}$ = $(d_{n-1} \cap f_{n-1}) \rightarrow'
d_n$. In general, without pseudo-complement, there is no
\textit{unique best} $e$ that maximises precision (see example
5.1). The construction is generalised for $g = p_1(\vec{x}_1),
\ldots, p_n(\vec{x}_n)$, by putting $e_n = \mathcal{C}$ and $e_i =
d_{i} \cap ((d_{i} \cap f_{i}) \rightarrow' e_{i+1})$ = $d_i \cap
(f_{i} \rightarrow' e_{i+1})$ for $1 \leq i < n$. Then $\langle g;
c \rangle \not\rreduce^{\star}_{P} \lozenge$ if $c \in e_1$ as
required. This iterated application of $\rightarrow'$ to propagate
requirements right-to-left is the very essence of the backward
analysis.

\begin{example}\label{exam-notunique} \rm
Returning to examples~3.2--3.5, let $\alpha_{EPos}(C) = \curlyvee
\{ \alpha(\theta) \mid \theta \in mgu(E) \wedge E \in C \}$,
$\gamma_{EPos}(f) = \cup \{ C \in \wp^{\downarrow}(Eqn) \mid
\alpha_{EPos}(C) \models f \}$ and $\rho_{EPos} = \gamma_{EPos}
\circ \alpha_{EPos}$. Note that $C_1 \cap C_2 = \gamma_{EPos}(f_1
\wedge f_2)$ and $C_1 \myvee' C_2 = \gamma_{EPos}(f_1 \curlyvee
f_2)$ where $C_i = \gamma_{EPos}(f_i)$. By defining $\exists'$ and
$d'$ in an analogous way to example~3.5, a semi-morphism
$\rho_{EPos}$ is constructed between $\langle
\wp^{\downarrow}(Eqn), \subseteq, \cup, \cap, Eqn, \emptyset,
\exists, d \rangle$ and $\langle
\rho_{EPos}(\wp^{\mydownarrow}(Eqn)), \subseteq, \myvee', \cap,
Eqn, \rho_{EPos}(\emptyset), \exists', d' \rangle$. Recall that
$\rho_{EPos}(\wp^{\mydownarrow}(Eqn))$ is \textit{not} a cHa. Now
consider the problem of inferring an initial $c$ for $\langle
p_{n-1}(\vec{x}_{n-1}), p_n(\vec{x}_n); c \rangle$ within
$\rho_{EPos}(\wp^{\mydownarrow}(Eqn))$. In particular let $d_{n-1}
= \gamma_{EPos}(1)$, $f_{n-1}$ = \mbox{$\gamma_{EPos}(x
\Leftrightarrow y)$} and $d_n = \gamma_{EPos}(x \wedge y)$. Then
$e_j \cap (d_{n-1} \cap f_{n-1}) \subseteq d_n$ for $e_1 =
\gamma_{EPos}(x)$ and $e_2 = \gamma_{EPos}(y)$ but $(e_1 \myvee'
e_2) \cap (d_{n-1} \cap f_{n-1}) = \gamma_{EPos}((x \curlyvee y)
\wedge 1 \wedge (x \Leftrightarrow y)) = \gamma_{EPos}(x
\Leftrightarrow y) \not\subseteq \gamma_{EPos}(x \wedge y) = d_n$.
Thus there is no unique $e$ maximising precision.
\end{example}

\begin{example} \rm Identity $\rho_{\mathrm{id}} = \lambda x . x$
is the trivial semi-morphism between $\langle
\wp^{\downarrow}(\mathcal{C}), \subseteq, \cup, \cap, \mathcal{C},
\emptyset, \exists, d \rangle$ and $\langle
\wp^{\downarrow}(\mathcal{C}), \subseteq, \cup, \cap, \mathcal{C},
\emptyset, \exists, d \rangle$ where the pseudo-complement is
given by $C_1 \rightarrow' C_2 = \{ c \in \mathcal{C} \mid \forall
c' \mymodels c \, . \, c' \in C_1 \Rightarrow c' \in C_2 \}$
\cite{B67}.
\end{example}

\begin{example} \rm Recall that
$\rho_{Pos}$ is a semi-morphism between $\langle
\wp^{\downarrow}(Eqn), \subseteq, \cup, \cap, Eqn, \emptyset,
\exists, d \rangle$ and $\langle
\rho_{Pos}(\wp^{\mydownarrow}(Eqn)), \subseteq, \myvee', \cap, Eqn,
\rho_{Pos}(\emptyset), \exists', d' \rangle$. Although
$\myvee' \neq \cup$,
$\rho_{Pos}(\wp^{\mydownarrow}(Eqn))$ is a sub-cHa of
$\wp^{\downarrow}(Eqn)$ with respect to $\cap$ and $\rightarrow'$ \cite{S97}.
Moreover, pseudo-complement (intuitionistic implication) $\rightarrow'$
coincides with classic implication $\Rightarrow$ in the sense
that $C_1 \rightarrow' C_2 = \gamma_{Pos}(f_1 \Rightarrow f_2)$ where
$C_i = \gamma_{Pos}(f_i)$. This follows since
$V \models f_2 \models (\neg f_1) \vee f_2$ and thus $f_1 \Rightarrow f_2 \in Pos$.
Moreover,
$f_1 \wedge f \models f_2$ iff $\models (f_1 \wedge f) \Rightarrow
f_2$ iff $\models f \Rightarrow (\neg f_1) \vee f_2$ iff $f
\models (\neg f_1) \vee f_2$.  Hence $C_1 \rightarrow' C_2 =
\myvee' \{ C \in \rho_{Pos}(\wp^{\mydownarrow}(Eqn)) \mid
C_1 \cap C \subseteq C_2 \} = \gamma_{Pos}(f_1 \Rightarrow f_2)$.
Thus $\rightarrow'$ is finitely
computable for $\rho_{Pos}$. Finally note that $\neg$ and $\vee$ are
defined on $Bool$ rather than $Pos$ since $\neg f \not\in Pos$ iff
$f \in Pos$.
\end{example}

\begin{example} \rm
Now consider the problem of inferring an initial $c$ for $\langle
p_{n-1}(\vec{x}_{n-1}), p_n(\vec{x}_n); c \rangle$ within
$\rho_{Pos}(\wp^{\mydownarrow}(Eqn))$. Analogous to
example~\ref{exam-notunique}, let $d_{n-1} = \gamma_{Pos}(1)$,
$f_{n-1}$ = \mbox{$\gamma_{Pos}(x \Leftrightarrow y)$} and $d_n =
\gamma_{Pos}(x \wedge y)$. Then $e_j \cap (d_{n-1} \cap f_{n-1})
\subseteq d_n$ for $e_1 = \gamma_{Pos}(x)$ and $e_2 =
\gamma_{Pos}(y)$ and $(e_1 \myvee' e_2) \cap (d_{n-1} \cap
f_{n-1}) = \gamma_{Pos}((x \vee y) \wedge 1 \wedge (x
\Leftrightarrow y)) = \gamma_{Pos}(x \wedge y) = d_n$. Thus there
is a unique $e$ maximising precision.
\end{example}

Since \mbox{$\langle \rho(\wp^{\mydownarrow}(\mathcal{C})),
\subseteq, \myvee', \cap, \mathcal{C}, \rho(\emptyset), \exists',
d' \rangle$} is a cylindric constraint system, it follows that $e
\subseteq \exists'_x(e)$ for all $e \in
\rho(\wp^{\mydownarrow}(\mathcal{C}))$. A consequence of $e
\subseteq \exists'_x(e)$ is that projection approximates from
above. Approximation from above, however, is not entirely
appropriate for backward analysis. In particular, observe that if
$\langle g; c \rangle \not\rreduce^{\star}_{P} \lozenge$ for all
$c \in e$, then it does not necessarily follow that $\langle g; c
\rangle \not\rreduce^{\star}_{P} \lozenge$ for all $c \in
\exists'_x(e)$. What is required is a dual notion of projection,
say denoted $\forall'$, that approximates from below. Then
$\langle g; c \rangle \not\rreduce^{\star}_{P} \lozenge$ for all
$c \in \forall'_x(e)$. Although $\forall'$ is an abstract
operator, the concept is defined for an arbitrary cylindric
constraint system for generality.

\begin{definition} \rm If $\langle \mathcal{C}, \mymodels, \myvee, \mywedge, 1,
0, \exists, d \rangle$ is a cylindric constraint system and $x \in
V$ then $\forall_x : \mathcal{C} \rightarrow \mathcal{C}$ is a
monotonic operator such that: $\exists_x(\forall_x(c)) \mymodels
c$ and $c \mymodels \forall_x(\exists_x(c))$ for all $c \in
\mathcal{C}$.
\end{definition}

\noindent Recall that $\exists_x$ is monotonic and thus $\alpha$
is the lower adjoint of $\gamma$ and $\gamma$ is the upper adjoint
of $\alpha$. More exactly, it follows that $\forall_x$ can be
automatically constructed from $\exists_x$ by $\forall_x(c) =
\myvee \{ c' \in \mathcal{C} \mid \exists_x(c') \mymodels c \}$.
Observe that this ensures that $\forall_x$ is the most precise
projection operator from below. For succinctness, define
$\forall_{\{ x_1, \ldots, x_n \} }(c)$ = $\forall_{x_1}( \ldots
(\forall_{x_n}(c)))$ and $\overline{\forall}_{X}(c)$ =
$\forall_{FV(c) \setminus X}(c)$.

\begin{example} \rm For $\rho_{\mathrm{id}}$,
let $\forall'_x(C) = \mydownarrow(\{ c \in C \mid \exists_x(c) = c
\})$.
\end{example}

\begin{example} \rm For $\rho_{Pos}$,
let $\forall'_x(C) = \gamma_{Pos}(f')$ if $f' \in Pos$ otherwise
$\forall'_x(C) = \gamma_{Pos}(0)$ where $C = \gamma_{Pos}(f)$ and
$f' = f[x \mapsto 0] \wedge f[x \mapsto 1]$. Observe that
$\exists_x(f)[x \mapsto 0] \wedge \exists_x(f)[x \mapsto 1]$ =
$\exists_x(f)$ and hence $C \subseteq \exists'_x(C) =
\forall'_x(\exists'_x(C))$ as required. Moreover, if
$\forall'_x(C) = \gamma_{Pos}(0)$ then $\exists'_x(\forall'_x(C))
= \gamma_{Pos}(0) \subseteq C$. Otherwise $\exists_x(f[x \mapsto
0] \wedge f[x \mapsto 1]) = f[x \mapsto 0] \wedge f[x \mapsto 1]
\models f$. Thus $\exists'_x(\forall'_x(C)) \subseteq C$ as
required. Finally, note that $\forall'_x$ is finitely computable
for $\rho_{Pos}$. For example if $C_i = \gamma_{Pos}(f_i)$, $f_1 =
(x \Leftarrow y)$, $f_2 = (x \wedge y)$ and $f_3 = (x \vee y)$,
then $\forall'_x(C_i) = \gamma_{Pos}(f_i')$ where $f'_1 = 0$,
$f'_2 = 0$ and $f'_3 = y$.
\end{example}

\noindent Backward analysis can now be formalised as
follows.

\begin{definition} \rm
Given a constraint logic program $P$ over a semi-cylindric
constraint system $\mathcal{C}$ with assertions over
$\rho(\wp^{\mydownarrow}(\mathcal{C}))$, the operator
$\mathcal{D}^{\rho, \mathcal{C}}_P : Int^{\mathrm{cod}(\rho)}
\rightarrow Int^{\mathrm{cod}(\rho)}$ is defined by:
\[
\mathcal{D}^{\rho, \mathcal{C}}_P(D) = \bigcup \left\{ E \left| \,
\begin{array}{@{}ccc@{}}
\forall & [{p(\vec{x}) \neck e}]_{\approx} \in E & . \\
\forall & p(\vec{x}) \neck C \separate c, p_1(\vec{x}_1), \ldots, p_n(\vec{x}_n) \in P & . \\
\exists & \{ [{p_i(\vec{x}_i) \neck f_i}]_{\approx} \}_{i = 1}^{n} \subseteq F & . \\
\exists & \{ [{p_i(\vec{x}_i) \neck d_i}]_{\approx} \}_{i = 1}^{n} \subseteq D & . \\
& e_{n+1} = \mathcal{C} \; \wedge \; e_{i} = d_i \cap (f_i \rightarrow' e_{i+1}) & \wedge \\
& e \subseteq \overline{\forall}'_{\vec{x}}(e_0) \; \wedge \; e_0 = C \cap (\rho(\mydownarrow(c)) \rightarrow' e_1) & \\
\end{array}
\right. \right\}
\]
where $[F]_{\equiv} =
\mathcal{F}^\mathcal{\mathrm{cod}(\rho)}(\rho(\mydownarrow(\mathrm{CLP}(P))))$.
\end{definition}

\noindent Since $\mathcal{D}$ is parameterised by $\rho$ and
$\mathcal{C}$ it can interpreted as a backward analysis framework.
$\mathcal{D}$ requires $F$, the success patterns of the program
obtained by discarding the assertions, to be pre-computed.
$\mathcal{D}$ considers each clause in the program in turn and
calculates those states which ensure that the clause (and those it
calls) will not violate an assertion. An abstraction which
characterises these states is calculated by propagating
requirements, represented as abstractions, right-to-left by
repeated application of pseudo-complement. Projection from below
then computes those states which, when restricted to the head
variables, still ensure that no error arises in the clause (and
those it calls). Repeated application of $\mathcal{D}$ yields a
decreasing sequence of interpretations.

The operator $\mathcal{D}^{\rho, \mathcal{C}}_P$ lifts to
$Int^{\mathcal{\mathrm{cod}(\rho)}}/\!\!\equiv$ by
$\mathcal{D}^{\rho, \mathcal{C}}_P([D]_{\equiv}) =
[\mathcal{D}^{\rho, \mathcal{C}}_P(D)]_{\equiv}$. Since \linebreak
\mbox{$\langle \mathrm{cod}(\rho), \sqsubseteq, \sqcup, \sqcap
\rangle$} is a complete lattice, $\mathcal{D}^{\rho,
\mathcal{C}}_P$ will possess a $\mathrm{gfp}$ if
$\mathcal{D}^{\rho, \mathcal{C}}_P$ is monotonic. The existence of
$\mathrm{gfp}(\mathcal{D}^{\rho, \mathcal{C}}_P)$ is guaranteed by
the following result since co-continuity implies monotonicity.

\begin{proposition}\label{prop-coco} \rm
$\mathcal{D}^{\rho, \mathcal{C}}_P :
Int^{\mathrm{cod}(\rho)}/\!\!\equiv \; \rightarrow
Int^{\mathrm{cod}(\rho)}/\!\!\equiv$ is co-continuous.
\end{proposition}

\noindent Since $\mathrm{gfp}(\mathcal{D}^{\rho, \mathcal{C}}_P)$
exists, a backward fixpoint semantics can be defined \linebreak
\mbox{$\mathcal{D}^{\rho, \mathcal{C}}(P)$ =
$\mathrm{gfp}(\mathcal{D}^{\rho, \mathcal{C}}_P)$} and computed by
lower Kleene iteration. To establish a connection between
$\mathcal{D}^{\rho, \mathcal{C}}(P)$ and the operational semantics
of $P$, it is useful to annotate the goals of a state with their
depth in the computation tree. To formalise this idea $\rreduce_P$
is lifted to the annotated states $Conf_{\lozenge} = Conf \cup \{
\lozenge \}$ where $Conf = Goal \times \mathcal{C} \times
\nat^{\star}$ to obtain the transition system $\rrreduce_P$.

\begin{definition} \rm Given a constraint logic program with assertions $P$ over a
semi-cylindric constraint system $\mathcal{C}$,
\mbox{$\rrreduce_{P} \subseteq Conf \times Conf_{\lozenge}$} is
the least relation such that:
\[
\langle p(\vec{x}), g; c; n \cdot h \rangle \rrreduce_{P} \left\{
\begin{array}{l@{\;}l}

\lozenge & \mathrm{if} \; \langle p(\vec{x}), g; c \rangle
\rreduce_{P} \lozenge \\

\langle g', g; c'; (n+1)^{|g'|} \cdot h \rangle & \mathrm{if} \;
\langle p(\vec{x}), g; c \rangle \rreduce_{P} \langle g', g; c'
\rangle

\end{array} \right.
\]
\end{definition}

\noindent The sequence $(n+1)^{|g'|}$ denotes $|g'|$
concatenations of $n+1$. The following result relates the depth of
the goals of the annotated states to the iterates obtained by
lower Kleene iteration. Informally, it says that if a constrained
atom $p(\vec{x}) \neck e$ occurs in the interpretation obtained by
applying $\mathcal{D}$ $k$ times, and $e$ characterises an initial
state (in a certain sense), and the depth of the goals in a
derivation starting at the initial state does not exceed $k$, then
the derivation will not violate an assertion. The main safety
theorem flows out of this result.

\begin{lemma}\label{lemma-demand} \rm
Let $\langle p(\vec{y}); c''; 1 \rangle = s_1
\rrreduce_{P}^{\star} s_n \rrreduce_{P} \lozenge$, $s_i = \langle
g_i; c_i; h_i \rangle$ and $(\mathcal{D}^{\rho,
\mathcal{C}}_{P})^k(\top) = [D_k]_{\equiv}$. \linebreak If
$\max(\{ \max(h_i) \mid 1 \leq i \leq n \}) \leq k$ and
\mbox{$[p(\vec{y}) \neck e]_{\approx} \in D_k$} then
$\overline{\exists}_{\vec{y}}(c'') \not\in
\overline{\exists}_{\vec{y}}(e)$.
\end{lemma}

\begin{theorem}\label{theorem-demand} \rm
If $\mathcal{D}^{\rho, \mathcal{C}}(P) = [D]_{\equiv}$,
$[p(\vec{y}) \neck e]_{\approx} \in D$ and $c \in
\overline{\exists}_{\vec{y}}(e)$ then $\langle p(\vec{y}); c
\rangle \not\rreduce_{P}^{\star} \lozenge$.
\end{theorem}

\section{Experimental evaluation}\label{sect-implement}

In order to evaluate the usefulness of the analysis framework
presented in section~\ref{sect-collect}, a backward $Pos$ analyser
has been constructed for inferring calling modes. The fixpoint
component of the analyser is coded in SICStus Prolog 3.8.3. The
domain operations are coded in C and are essentially the binary
decision diagram (BDD) routines written by Armstrong and Schachte
\cite{AMSS98}. The analyser takes, as input, a program written in
a declarative subset of ISO~Prolog. It outputs a mode for each
program predicate. The safety result of
theorem~\ref{theorem-demand} ensures that if a call to a predicate
is at least as instantiated as the inferred mode, then the call
will not violate an instantiation requirement. Modes are expressed
as grounding dependencies \cite{AMSS98}.

The implementation follows the framework defined in
section~\ref{sect-collect} very closely. The analyser was
straightforward to implement as it is essentially two bottom-up
fixpoint computations: one for $\mathcal{F}$ and the other for
$\mathcal{D}$. The only subtlety is in handling the builtins. For
each builtin, it is necessary to select a grounding dependency
that is sufficient for avoiding an instantiation error. This is an
lower approximation (the \textit{required mode} of
table~\ref{table-abstract}). It is also necessary to specify
behaviour on success. This is an upper-approximation (the
\textit{success mode} of table~\ref{table-abstract}). The lower
approximations are the assertions that are added to Prolog program
to obtain a constraint logic program with assertions.

Interestingly, the success mode does not always entail the
required mode. Univ (=..) illustrates this. A sufficient but not
necessary condition for univ not to error is that either the first
or second argument is ground. This cannot be weakened in $Pos$
(but could be weakened in a type dependency domain \cite{CL00}
that expressed rigid lists). The success mode is that the first
argument is ground iff the second argument is ground (which does
not entail the required mode). Note too that keysort and sort
error if their first argument is free. A sufficient mode for
expressing this requirement is that the first argument is ground.
Again, this requirement cannot be weakened in $Pos$.

\begin{table}
\begin{tabular}{@{}p{2.5in}|c|c@{}}
\textit{builtin} & \textit{required mode} & \textit{success mode} \\

\cline{1-3} \mbox{$t_1$ == $t_2$}, \mbox{$t_1$ $\backslash$==
$t_2$}, \mbox{$t_1$ @$<$ $t_2$}, \mbox{$t_1$ @$>$ $t_2$},
\mbox{$t_1$ @=$<$ $t_2$}, \mbox{$t_1$ @$>$= $t_2$}, \mbox{$t_1$
$\backslash$= $t_2$}, !, compound($t_1$), display($t_1$), listing,
listing($t_1$), nl, nonvar($t_1$), print($t_1$),
\mbox{portray\_clause($t_1$)}, read($t_1$), repeat, true,
var($t_1$), write($t_1$), writeq($t_1$) & $true$ & $true$ \\

\cline{1-3} atom($t_1$), atomic($t_1$), \mbox{compare($t_1, t_2,
t_3$)}, float($t_1$), ground($t_1$), integer($t_1$),
number($t_1$) & $true$ & $f_1$ \\

\cline{1-3} \mbox{length($t_1, t_2$)} & $true$ & $f_2$ \\

\cline{1-3} \mbox{statistics($t_1, t_2$)} & $true$ & $g_1$ \\


\cline{1-3} \cline{1-3} abort, fail, false & $true$ & $false$ \\

\cline{1-3} \mbox{keysort($t_1, t_2$)}, \mbox{sort($t_1, t_2$)} & $f_1$ & $g_2$ \\

\cline{1-3} tab($t_1$), put($t_1$) & $f_1$ & $f_1$ \\

\cline{1-3} \mbox{$t_1$ is $t_2$} & $f_2$ & $g_1$ \\

\cline{1-3} \mbox{$t_1$ =:= $t_2$}, \mbox{$t_1$ =$\backslash$=
$t_2$}, \mbox{$t_1 < t_2$}, \mbox{$t_1 > t_2$},
\mbox{$t_1$ =$<$ $t_2$}, \mbox{$t_1$ $>$= $t_2$} & $g_1$ & $g_1$ \\

\cline{1-3} \mbox{arg($t_1$, $t_2$, $t_3$)} & $g_1$ & $g_3$ \\

\cline{1-3} \mbox{name($t_1, t_2$)} & $g_4$ & $g_1$ \\

\cline{1-3} \mbox{$t_1$ =.. $t_2$} & $g_4$ & $g_2$ \\

\cline{1-3} \mbox{functor($t_1, t_2, t_3$)} & $g_5$ & $g_{6}$

\end{tabular}
\caption{\label{table-abstract} Abstracting builtins where
\mbox{$f_i = \wedge \mathrm{var}(t_i)$}, \mbox{$g_1$ = $f_1 \wedge
f_2$},
\mbox{$g_2 = f_1 \Leftrightarrow f_2$}, \mbox{$g_3 = f_1 \wedge
(f_2 \Rightarrow f_3)$}, \mbox{$g_4 = f_1 \vee f_2$}, \mbox{$g_5 =
f_1 \vee (f_2 \wedge f_3)$} and \mbox{$g_{6} = f_2 \wedge f_3$}.}
\end{table}

The analyser has been applied to some standard Prolog benchmarks
which can be found at
http://www.oakland.edu/\verb+~+l2lu/benchmarks-BG.zip. The results
of the analysis, that is, the calling modes for the predicates in
the smaller benchmarks, are given in table~\ref{table-precision}.
The results, though surprising in some cases (see sort of permSort
and insert of treesort for example) have been verified by hand and
appear to be optimal for $Pos$. The analysis, of course, can be
applied to larger programs (though it becomes very difficult to
verify the results by hand) and table~\ref{table-speed}
demonstrates that the analysis scales smoothly to medium-scale
programs at least. The table lists the larger benchmarks (which
possibly include some unreachable code) in terms of increasing
\textit{size} measured by the total number of atoms in the source.
The \textit{abs} column records the time in milliseconds required
to read, parse and normalise the source into the ground program
representation used by the analyser; \textit{lfp} is the time
needed to compute the fixpoint characterising the success modes;
\textit{gfp} is the time needed to compute the calling modes; and
finally \textit{sum} is the total analysis time. This includes the
(usually negligible) overhead of annotating the source with the
modes required by builtins. Timings were performed on a Dell GX200
1GHz~PC with 128~MB memory running Windows~2000. The timings
suggest that the analysis is practical at least for medium-scale
programs (though the running time for BDDs can be sensitive to the
particular dependencies that arise). Moreover, with a
state-of-the-art GER factorised BDD package \cite{BS98} the
analysis would be faster. Interestingly, the time to compute the
lfp often dominates the whole analysis. BDD widening will be
required to analyse very large applications but this is a study
within itself \cite{HACK00}.

\begin{table}\centering
\begin{tabular}{l|l|c}
\textit{benchmark} & \textit{predicate} & \textit{mode} \\

\cline{1-3}
bubblesort & sort($x_1,x_2$) & $x_1$ \\
      & ordered($x_1$) & $x_1$ \\
      & append($x_1,x_2,x_3$) & $true$ \\

\cline{1-3}
dnf   & go & $true$ \\
    & dnf($x_1,x_2$) & $true$ \\
    & norm($x_1,x_2$) & $true$ \\
    & literal($x_1$) & $true$ \\

\cline{1-3}
heapify & greater($x_1,x_2$) & $x_1 \land x_2$ \\
    & adjust($x_1,x_2,x_3,x_4$) & $\left( \begin{array}{c@{\;}c}
                                    (x_1 \land x_4) & \lor \\
                                    (x_1 \land x_2 \land x_3) & \lor \\
                                    \multicolumn{2}{l}{(\neg x_2 \land \neg x_3 \land x_4)}
                                    \end{array} \right)$ \\
    & heapify($x_1,x_2$) & $x_1$ \\

\cline{1-3}
permSort  & select($x_1,x_2,x_3$) & $true$ \\
      & ordered($x_1$) & $x_1$ \\
      & permutation($x_1,x_2$) & $true$ \\
      & sort($x_1,x_2$) & $x_1\lor x_2$ \\

\cline{1-3}
queens   & noattack($x_1,x_2,x_3$) & $x_1 \land x_2 \land x_3$ \\
      & safe($x_1$) & $x_1$ \\
      & delete($x_1,x_2,x_3$) & $true$ \\
      & perm($x_1,x_2$) & $true$ \\
      & queens($x_1,x_2$) & $x_1\lor x_2$ \\

\cline{1-3}
quicksort  & append($x_1,x_2,x_3$) & $true$ \\
      & qsort($x_1,x_2$) & $x_1$ \\
      & partition($x_1,x_2,x_3,x_4$) & $x_2 \land (x_1 \lor (x_3 \land x_4$)$)$ \\

\cline{1-3}
treeorder  & member($x_1,x_2$) & $true$ \\
      & select($x_1,x_2,x_3$) & $true$ \\
      & split($x_1,x_2,x_3,x_4$) & $true$ \\
      & split($x_1, \ldots ,x_7$) & $true$ \\
      & visits2tree($x_1,x_2,x_3$) & $true$ \\
      & v2t($x_1,x_2,x_3$) & $true$ \\

\cline{1-3}
treesort  & tree\_to\_list\_aux($x_1,x_2,x_3$) & $true$ \\
      & tree\_to\_list($x_1,x_2$) & $true$ \\
      & list\_to\_tree($x_1,x_2$) & $x_1$ \\
      & insert\_list($x_1,x_2,x_3$) & $x_1 \land x_2$ \\
      & insert($x_1,x_2,x_3$) & $x_1 \land (x_2\lor x_3)$ \\
      & treesort($x_1,x_2$) & $x_1$
\end{tabular}
\caption{\label{table-precision} Precision of the Mode Analysis
(small benchmarks)}
\end{table}

\begin{table}
\begin{tabular}{@{}r|@{}r|@{}r@{}r@{}r@{}r@{}r|@{}r|@{}r@{}r@{}r@{}r@{}}
{\it file} & {\it size} & {\it abs} & {\it lfp} & {\it gfp} & {\it sum} &
{\it file} & {\it size} & {\it abs} & {\it lfp} & {\it gfp} & {\it sum} \\
\cline{1-12}
astar&100&10&10&0&20 &
tictactoe&258&20&10&10&40\\
fft&104&20&0&10&30 &
jons2&261&20&10&0&30\\
knight&105&10&0&0&10 &
kalah&269&30&10&20&60\\
browse\_wamcc&106&10&0&0&10 &
draw&289&70&91&40&201\\
cal\_wamcc&108&10&10&0&20 &
cs\_r&311&40&20&10&70\\

life&110&10&10&10&30 &
reducer&320&40&30&0&70\\
crypt\_wamcc&113&10&0&0&10 &
sdda&336&20&21&0&41\\
cry\_mult&118&10&10&10&30 &
bryant&349&30&120&21&171\\
browse&125&10&10&0&20 &
ga&363&50&30&20&100\\
bid&128&10&10&0&20 &
neural&378&30&10&0&40\\

disj\_r&148&30&0&10&40 &
press&381&30&20&0&50\\
consultant&151&20&0&10&30 &
peep&414&50&20&10&80\\
ncDP&156&10&10&0&20 &
nbody&421&40&20&20&80\\
tsp&162&30&20&10&60 &
eliza&432&50&20&0&70\\
elex\_scanner&165&20&10&0&30 &
read&434&40&20&10&70\\

robot&165&10&10&0&20 &
simple\_analyzer&512&90&701&20&811\\
sorts&172&0&10&10&20 &
ann&547&50&30&10&90\\
cs2&175&30&10&10&50 &
diffsimpsv&681&61&100&0&161\\
scc&175&10&141&0&151 &
arch1&692&50&40&10&100\\
bp0-6&201&20&10&0&30 &
asm&800&60&40&30&130\\

bnet&205&20&20&0&40 &
poker&962&81&70&10&161\\
jons&222&40&0&10&50 &
pentomino&981&50&40&80&170\\
mathlib&226&10&10&0&20 &
chat&1037&411&1422&1082&2915\\
intervals&230&20&10&10&40 &
sim\_v5-2&1308&80&70&0&150\\
barnes\_hut&240&40&30&40&110 &
semigroup&2328&180&90&60&350
\end{tabular}
\caption{\label{table-speed} Speed of the Mode Analysis
(medium-scale benchmarks)}\end{table}

\section{Related work}\label{sect-relate}

Our work was motivated by the recent revival of interest in logic
programming with assertions \cite{BWM97,PBH00a}. For example,
\cite{PBH00b} argues that it is useful to trap an unexpected call
to a predicate with an assertion otherwise a program may error at
a point that is far from the source of the problem. Moreover,
\cite{PBH00a} observe that predicates are normally written with an
expectation on the initial calling pattern, and hence provide an
\texttt{entry} assertion to make the, moding say, of the top-level
queries explicit. Our work shows how \texttt{entry} assertions can
be automatically synthesised which ensure that instantiation
errors do not occur while executing the program.

The most closely related work concerns the demand analysis of ccp
\cite{D93,FHW00}. A demand analysis for the ccp language Janus
\cite{SKL90} is proposed in \cite{D93} which determines whether or
not a predicate is uni-modal. A predicate is uni-modal iff the
argument tuple for each clause share the same minimal pattern of
instantiation necessary for reduction. The demand analysis of a
predicate simply traverses the head and guard of each clause to
determine the extent to which arguments have to be instantiated.
Body atoms need not be considered so the analysis does not involve
a fixpoint computation. A related paper \cite{DGB96} presents a
goal-dependent (forward) analysis that detects those ccp
predicates which can be scheduled left-to-right without deadlock.
If assertions are used to approximate synchronisation, then the
analysis described in this paper can be re-interpreted as a
backward suspension analysis of ccp under left-to-right
scheduling.

When reasoning about module interaction it can be advantageous to
reverse the traditional deductive approach to abstract
interpretation that is based on the abstract unfolding of abstract
goals. In particular \cite{Giaco98} shows how abduction and
abstraction can be combined to compute those properties that one
module must satisfy to ensure that its composition with another
fulfils certain requirements. Abductive analysis can, for
example, determine how an optimisation in one module depends on a
predicate defined in another module. Abductive analysis is related
to the backward analysis presented in this paper since abduction
is the inverse image of a forward semantics whereas
pseudo-complement is the inverse image of conjunction -- the basic
computational step in forward (and backward) semantics.

The termination inference engine of \cite{GC01} decomposes the cTI
analyser of \cite{Mesnard96b} into two components: a termination
checker \cite{CT99} and the backward analysis described in this
paper.  First, the termination inference engine computes a set of
binary clauses which describe possible loops in the program with
size relations. Second, a Boolean function is inferred for each
predicate that describes moding conditions sufficient for each
loop to only be executed a finite number of times. Third, the
backward analysis described in this paper is applied to infer
initial modes by calculating a greatest fixpoint which guarantee
that the moding conditions hold and thereby assure termination.
Interestingly, the cTI analyser involves a $\mu$-calculus solver
to compute the greatest fixpoint of an equivalent (though more
complex) system of equations. This seems to suggest that greatest
fixpoints are important in backward analysis.

Cousot and Cousot \cite{CC92} explain how a backward collecting
semantics can be deployed to precisely characterise states that
arise in finite SLD-derivations. First, they present a forward
collecting semantics that records the descendant states that arise
from a set of initial states. Second, they present a dual
(backward) collecting semantics that records those states which
occur as ascendant states of the final states. By combining both
semantics, they characterise the set of descendant states of the
initial states which are also ascendant states of the final states
of the transition system. This use of backward analysis is
primarily as a device to improve the precision of a classic
goal-dependent analysis. Our work is more radical in the sense
that it shows how a bottom-up analysis performed in a backward
fashion, can be used to characterise initial queries. Moreover it
is used for lower approximation rather than upper approximation.

Mazur, Janssens and Bruynooghe \cite{MJB2000} present a kind of
\textit{ad hoc} backward analysis to derive reuse conditions from
a goal-independent reuse analysis for Mercury \cite{mercury}. The
analysis propagates reuse information from a point where a
structure is decomposed in a clause to the point where the clause
is invoked in its parent clause. This is similar in spirit to how
demand is passed from a callee to a caller in the backward
analysis described in this paper. However, the reuse analysis does
not propagate information right-to-left across a clause using
pseudo-complement, and so one interesting topic for future work
will to be relate these two analyses. Another matter for future
work, will be to investigate the extent to which our backward mode
analysis can be reconstructed by inverting abstract functions
\cite{HL94}.

\section{Conclusion}\label{sect-conclude}

We have shown how abstract interpretation, and specifically a
backward analysis, can infer moding properties which if satisfied
by the initial query, come with the guarantee that the program and
query cannot generate instantiation errors. Backward analysis has
other applications in termination inference and also in inferring
queries for which the builtins called from within the program
behave predictably in the presence of rational trees. The analysis
is composed of two bottom-up fixpoint calculations, a lfp and a
gfp, both of which are straightforward to implement. The lfp
characterises success patterns. The gfp, uses these success
patterns to infer safe initial calling patterns. It propagates
moding requirements right-to-left, against the control-flow, using
the pseudo-complement operator. This operator fits with backward
analysis since it enables moding requirements to be minimised
(maximally weakened) in right-to-left propagation. This operator,
however, requires that the computational domain be closed under
Heyting completion (or equivalently condense). This requirement
seems reasonable because disjunctive dependencies occur frequently
in right-to-left propagation and therefore significant precision
would be lost if the requirement were relaxed. Experimental
evaluation has demonstrated that the analysis is practical in the
sense that it can infer calling modes for medium-scaled programs.
Finally, our work adds weight to the belief that condensing is an
important property in the analysis of logic programs.

\subsubsection*{Acknowledgements}
We thank Maurice Bruynooghe, Mike Codish, \mbox{Samir Genaim},
Roberto Giacobazzi, \mbox{Jacob Howe}, \mbox{Fred Mesnard},
Germ\'{a}n Puebla and Francesca \mbox{Scozzari} for helpful
discussions. We would also like to thank the anonymous referees
for their comments and Peter Schachte for his BDD analyser. We
also thank Roberto Bagnara for the use of some of the
\textsc{China} benchmarks. This work was supported, in part, by
EPSRC grant GR/MO8769.


\begin{thebibliography}{}

\bibitem[\protect\citename{Armstrong {\em et~al.}\relax, }1998]{AMSS98}
Armstrong, T., Marriott, K., Schachte, P., \& S{\o}ndergaard, H. (1998).
\newblock Two {C}lasses of {B}oolean {F}unctions for {D}ependency {A}nalysis.
\newblock {\em Science of {C}omputer {P}rogramming}, {\bf 31}(1), 3--45.

\bibitem[\protect\citename{Bagnara \& Schachte, }1999]{BS98}
Bagnara, R., \& Schachte, P. (1999).
\newblock Factorizing {E}quivalent {V}ariable {P}airs in {ROBDD}-{B}ased
  {I}mplementations of \textit{Pos}.
\newblock {\em Pages  471--485 of:} {\em {I}nternational {C}onference on
  {A}lgebraic {M}ethodology and {S}oftware {T}echnology}.
\newblock Lecture Notes in Computer Science, vol. 1548.
\newblock Springer-Verlag.

\bibitem[\protect\citename{Bagnara {\em et~al.}\relax, }2001]{BZGH01}
Bagnara, R., Zaffanella, E., Gori, R., \& Hill, P.~M. (2001).
\newblock Boolean {F}unctions for {F}inite-{T}ree {D}ependencies.
\newblock {\em Pages  575--589 of:} {\em International {C}onference on {L}ogic
  for {P}rogramming, {A}rtificial {I}ntelligence and {R}easoning}.
\newblock Lecture {N}otes in {A}rtificial {I}ntelligence, vol. 2250.
\newblock Springer-Verlag.

\bibitem[\protect\citename{Birkhoff, }1967]{B67}
Birkhoff, G. (1967).
\newblock {\em Lattice {T}heory}.
\newblock AMS Press.

\bibitem[\protect\citename{Bossi {\em et~al.}\relax, }1994]{BGLM94}
Bossi, A., Gabbrielli, M., Levi, G., \& Martelli, M. (1994).
\newblock {T}he s-{S}emantics {A}pproach: {T}heory and {A}pplications.
\newblock {\em The {J}ournal of {L}ogic {P}rogramming}, {\bf {19/20}},
  149--197.

\bibitem[\protect\citename{Boye {\em et~al.}\relax, }1997]{BWM97}
Boye, J., Drabent, W., \& Ma{\l}uszy{\'{n}}ski, J. (1997).
\newblock Declarative {D}iagnosis of {C}onstraint {P}rograms: an
  {A}ssertion-based {A}pproach.
\newblock {\em Pages  123--141 of:} {\em Proceedings of the {T}hird
  {I}nternational {W}orkshop on {A}utomated {D}ebugging}.
\newblock University of Link\"{o}ping Press.

\bibitem[\protect\citename{Codish \& Lagoon, }2000]{CL00}
Codish, M., \& Lagoon, V. (2000).
\newblock Type {D}ependencies for {L}ogic {P}rograms using {ACI}-unification.
\newblock {\em Theoretical {C}omputer {S}cience}, {\bf 238}, 131--159.

\bibitem[\protect\citename{Codish \& Taboch, }1999]{CT99}
Codish, M., \& Taboch, C. (1999).
\newblock A {S}emantic {B}asis for the {T}ermination {A}nalysis of {L}ogic
  {P}rograms.
\newblock {\em The {J}ournal of {L}ogic {P}rogramming}, {\bf 41}(1), 103--123.

\bibitem[\protect\citename{Cousot \& Cousot, }1982]{CC82}
Cousot, P., \& Cousot, R. (1982).
\newblock Inductive {P}rinciples for {P}roving {I}nvariance {P}roperties of
  {P}rograms.
\newblock {\em Pages  75--119 of:} {\em Tools and {N}otions for {P}rogram
  {C}onstruction}.
\newblock Cambridge University Press.

\bibitem[\protect\citename{Cousot \& Cousot, }1992]{CC92}
Cousot, P., \& Cousot, R. (1992).
\newblock Abstract {I}nterpretation and {A}pplication to {L}ogic {P}rograms.
\newblock {\em The {J}ournal of {L}ogic {P}rogramming}, {\bf 13}(2--3),
  103--179.

\bibitem[\protect\citename{Debray, }1993]{D93}
Debray, S. (1993).
\newblock {QD}-{J}anus: a {S}equential {I}mplementation of {J}anus in {P}rolog.
\newblock {\em {S}oftware {P}ractice and {E}xperience}, {\bf 23}(12),
  1337--1360.

\bibitem[\protect\citename{Debray {\em et~al.}\relax, }1992]{DGB96}
Debray, S., Gudeman, D., \& Bigot, P. (1992).
\newblock Detection and {O}ptimization of {S}uspension-free {L}ogic {P}rograms.
\newblock {\em The {J}ournal of {L}ogic {P}rogramming}, {\bf 29}(1--3),
  171--194.

\bibitem[\protect\citename{Drabent \& Ma{\l}uszy{\'{n}}ski, }1988]{DM88}
Drabent, W., \& Ma{\l}uszy{\'{n}}ski, J. (1988).
\newblock Inductive {A}ssertion {M}ethod for {L}ogic {P}rograms.
\newblock {\em Theoretical {C}omputer {S}cience}, {\bf 59}(1), 133--155.

\bibitem[\protect\citename{Dyber, }1991]{D91}
Dyber, P. (1991).
\newblock Inverse {I}mage {A}nalysis {G}eneralises {S}trictness {A}nalysis.
\newblock {\em Information and {C}omputation}, {\bf 90}(2), 194--216.

\bibitem[\protect\citename{Falaschi {\em et~al.}\relax, }2000]{FHW00}
Falaschi, M., Hicks, P., \& Winsborough, W. (2000).
\newblock Demand {T}ransformation {A}nalysis for {C}oncurrent {C}onstraint
  {P}rograms.
\newblock {\em The {J}ournal of {L}ogic {P}rogramming}, {\bf 41}(3), 185--215.

\bibitem[\protect\citename{Fil\'{e} \& Ranzato, }1994]{FR94}
Fil\'{e}, G., \& Ranzato, F. (1994).
\newblock Improving {A}bstract {I}nterpretations by {S}ystematic {L}ifting to
  the {P}owerset.
\newblock {\em Pages  655--669 of:} {\em International {L}ogic {P}rogramming
  {S}ymposium}.
\newblock MIT Press.

\bibitem[\protect\citename{Genaim \& Codish, }2001]{GC01}
Genaim, S., \& Codish, M. (2001).
\newblock {I}nferring {T}ermination {C}onditions for {L}ogic {P}rograms using
  {B}ackwards {A}nalysis.
\newblock {\em Pages  681--690 of:} {\em International {C}onference on {L}ogic
  for {P}rogramming, {A}rtificial {I}ntelligence and {R}easoning}.
\newblock Lecture {N}otes in {A}rtificial {I}ntelligence, vol. 2250.
\newblock Springer-Verlag.

\bibitem[\protect\citename{Giacobazzi, }1998]{Giaco98}
Giacobazzi, R. (1998).
\newblock Abductive {A}nalysis of {M}odular {L}ogic {P}rograms.
\newblock {\em Journal of {L}ogic and {C}omputation}, {\bf 8}(4), 457--484.

\bibitem[\protect\citename{Giacobazzi \& Scozzari, }1998]{GS98}
Giacobazzi, R., \& Scozzari, F. (1998).
\newblock A {L}ogical {M}odel for {R}elational {A}bstract {D}omains.
\newblock {\em {ACM} {T}ransactions on {P}rogramming {L}anguages and
  {S}ystems}, {\bf 20}(5), 1067--1109.

\bibitem[\protect\citename{Giacobazzi {\em et~al.}\relax, }1995]{GDL95}
Giacobazzi, R., Debray, S., \& Levi, G. (1995).
\newblock Generalized {S}emantics and {A}bstract {I}nterpretation for
  {C}onstraint {L}ogic {P}rograms.
\newblock {\em The {J}ournal of {L}ogic {P}rogramming}, {\bf 25}(3), 191--248.

\bibitem[\protect\citename{Giacobazzi {\em et~al.}\relax,
  }1998]{GiacobazziSR98}
Giacobazzi, R., Ranzato, F., \& Scozzari, F. (1998).
\newblock Building {C}omplete {A}bstract {I}nterpretations in a {L}inear
  {L}ogic-based {S}etting.
\newblock {\em Pages  215--229 of:} {\em Static {A}nalysis {S}ymposium}.
\newblock Lecture {N}otes in {C}omputer {S}cience, vol. 1503.
\newblock Springer-Verlag.

\bibitem[\protect\citename{Hall \& Wise, }1989]{HW89}
Hall, C., \& Wise, D. (1989).
\newblock Generating {F}unction {V}ersions with {R}ational {S}trictness
  {P}atterns.
\newblock {\em Science of {C}omputer {P}rogramming}, {\bf 12}, 39--74.

\bibitem[\protect\citename{Heaton {\em et~al.}\relax, }2000]{HACK00}
Heaton, A., Abo-Zaed, M., Codish, M., \& King, A. (2000).
\newblock A {S}imple {P}olynomial {G}roundness {A}nalysis for {L}ogic
  {P}rograms.
\newblock {\em The {J}ournal of {L}ogic {P}rogramming}, {\bf 45}(1--3),
  143--156.

\bibitem[\protect\citename{Hughes \& Launchbury, }1994]{HL94}
Hughes, R.~J.~M., \& Launchbury, J. (1994).
\newblock Reversing {A}bstract {I}nterpretations.
\newblock {\em Science of {C}omputer {P}rogramming}, {\bf 22}, 307--326.

\bibitem[\protect\citename{Jaffar \& Maher, }1994]{JM94}
Jaffar, J., \& Maher, M.~J. (1994).
\newblock Constraint {L}ogic {P}rogramming: {A} {S}urvey.
\newblock {\em {T}he {J}ournal of {L}ogic {P}rogramming}, {\bf {19/20}},
  503--581.

\bibitem[\protect\citename{Langen, }1991]{L91}
Langen, A. (1991).
\newblock {\em Advanced {T}echniques for {A}pproximating {V}ariable {A}liasing
  in {L}ogic {P}rograms}.
\newblock Ph.D. thesis, Computer {S}cience {D}epartment, Los {A}ngeles,
  {C}alifornia 90089-0782.

\bibitem[\protect\citename{Marriott \& S{\o}ndergaard, }1993]{MS93}
Marriott, K., \& S{\o}ndergaard, H. (1993).
\newblock Precise and {E}fficient {G}roundness {A}nalysis for {L}ogic
  {P}rograms.
\newblock {\em {ACM} {L}etters on {P}rogramming {L}anguages and {S}ystems},
  {\bf 2}(4), 181--196.

\bibitem[\protect\citename{Mazur {\em et~al.}\relax, }2000]{MJB2000}
Mazur, N., Janssens, G., \& Bruynooghe, M. (2000).
\newblock A {M}odule {B}ased {A}nalysis for {M}emory {R}euse in {M}ercury.
\newblock {\em Pages  1255--1269 of:} {\em Computational {L}ogic}.
\newblock Lecture {N}otes in {A}rtificial {I}ntelligence, vol. 1861.

\bibitem[\protect\citename{Mesnard, }1996]{Mesnard96b}
Mesnard, F. (1996).
\newblock Inferring {L}eft-terminating {C}lasses of {Q}ueries for {C}onstraint
  {L}ogic {P}rograms.
\newblock {\em Pages  7--21 of:} {\em Joint {I}nternational {C}onference and
  {S}ymposium on {L}ogic {P}rogramming}.
\newblock MIT Press.

\bibitem[\protect\citename{Mesnard \& Neumerkel, }2001]{Mesnard01}
Mesnard, F., \& Neumerkel, U. (2001).
\newblock Applying {S}tatic {A}nalysis {T}echniques for {I}nferring
  {T}ermination {C}onditions of {L}ogic {P}rograms.
\newblock {\em Pages  93--110 of:} {\em Static {A}nalysis {S}ymposium}.
\newblock Lecture Notes in Computer Science, vol. 2126.
\newblock Springer-{V}erlag.

\bibitem[\protect\citename{Mycroft, }1981]{M81}
Mycroft, A. (1981).
\newblock {\em Abstract {I}nterpretation and {O}ptimising {T}ransformations for
  {A}pplicative {P}rograms}.
\newblock Ph.D. thesis, University of {E}dinburgh.

\bibitem[\protect\citename{Puebla {\em et~al.}\relax, }2000a]{PBH00a}
Puebla, G., Bueno, F., \& Hermenegildo, M. (2000a).
\newblock An {A}ssertion {L}anguage for {C}onstraint {L}ogic {P}rograms.
\newblock {\em Pages  23--61 of:} {\em {A}nalysis and {V}isualization {T}ools
  for {C}onstraint {P}rogramming}.
\newblock Lecture Notes in Computer Science, vol. 1870.
\newblock Springer-Verlag.

\bibitem[\protect\citename{Puebla {\em et~al.}\relax, }2000b]{PBH00b}
Puebla, G., Bueno, F., \& Hermenegildo, M. (2000b).
\newblock A {G}eneric {P}reprocessor for {P}rogram {V}alidation and
  {D}ebugging.
\newblock {\em Pages  63--107 of:} {\em {A}nalysis and {V}isualization {T}ools
  for {C}onstraint {P}rogramming}.
\newblock Lecture Notes in Computer Science, vol. 1870.
\newblock Springer-Verlag.

\bibitem[\protect\citename{Saraswat {\em et~al.}\relax, }1990]{SKL90}
Saraswat, V., Kahn, K., \& Levy, J. (1990).
\newblock Janus: a {S}tep {T}owards {D}istributed {C}onstraint {P}rogramming.
\newblock {\em Pages  431--446 of:} {\em North {A}merican {C}onference on
  {L}ogic {P}rogramming}.
\newblock MIT Press.

\bibitem[\protect\citename{Scozzari, }{to appear}]{S97}
Scozzari, F. ({to appear}).
\newblock Logical {O}ptimality of {G}roundness {A}nalysis.
\newblock {\em Theoretical {C}omputer {S}cience}.

\bibitem[\protect\citename{Somogyi {\em et~al.}\relax, }1996]{mercury}
Somogyi, Z., Henderson, F., \& Conway, T. (1996).
\newblock The execution algorithm of {M}ercury, an efficient purely declarative
  logic programming language.
\newblock {\em {T}he {J}ournal of {L}ogic {P}rogramming}, {\bf 29}(1--3),
  17--64.

\bibitem[\protect\citename{{van Dalen}, }1997]{VanDalen}
{van Dalen}, D. (1997).
\newblock {\em Logic and {S}tructure}.
\newblock Springer.

\bibitem[\protect\citename{Wadler \& Hughes, }1987]{WH87}
Wadler, P., \& Hughes, R.~J.~M. (1987).
\newblock Projections for {S}trictness {A}nalysis.
\newblock {\em Pages  385--407 of:} {\em Functional {P}rogramming and
  {C}omputer {A}rchitecture}.
\newblock Lecture Notes in Computer Science, vol. 274.
\newblock Springer-Verlag.

\end{thebibliography}

\newpage

\appendix

\section{Proof appendix}\label{sect-appendix}


\begin{proof}[Proof for proposition~\ref{prop-down}] \rm Proof by
induction. Let $I_0 = \emptyset$, $I_0' = \emptyset$, $I_{k+1} =
\mathcal{F}^\mathcal{C}_{P}(I_k)$ and $I_{k+1}' =
\mathcal{F}^{\wp^{\mydownarrow}(\mathcal{C})}_{\mydownarrow(P)}(I_k')$.
To show $\mydownarrow(I_k) \sqsubseteq I_k'$ since then it follows
that $\mydownarrow(\mathrm{lfp}(\mathcal{F}^\mathcal{C}_{P}))$ =
$\mydownarrow(\sqcup_{k \in \nat} I_k )$ $\sqsubseteq$ $\sqcup_{k
\in \nat} \mydownarrow(I_k)$ $\sqsubseteq$ $\sqcup_{k \in \nat}
I_k'$ =
$\mathrm{lfp}(\mathcal{F}^{\wp^{\mydownarrow}(\mathcal{C})}_{\mydownarrow(P)})$.
The base case is trivial so suppose $\mydownarrow(I_k) \sqsubseteq
I_k'$. \linebreak Let \mbox{$[p(\vec{x}) \neck c']_{\approx} \in
I_{k+1}$}. Then there exists $p(\vec{x}) \neck c,
p_{1}(\vec{x}_1), \ldots, p_{n}(\vec{x}_n) \in P$ and \linebreak
\mbox{$\{ [{p_{i}(\vec{x}_i) \neck c_i}]_{\approx} \}_{i=1}^{n}
\subseteq I_k$} such that $c' = c \mywedge \mywedge_{i = 1}^{n}
{\overline{\exists}_{\vec{x_i}}}(c_i)$. Observe that
$\mydownarrow(c')$ $\subseteq$ \linebreak \mbox{$\mydownarrow(c)
\cap \cap_{i = 1}^{n}
\mydownarrow({\overline{\exists}_{\vec{x_i}}}(c_i))$} $\subseteq$
\mbox{$\mydownarrow(c) \cap \cap_{i = 1}^{n}
{\overline{\exists}_{\vec{x_i}}}(\mydownarrow(c_i))$}. But by the
inductive hypothesis, there exist \mbox{$\{ [{p_{i}(\vec{x}_i)
\neck c_i'}]_{\approx} \}_{i=1}^{n} \subseteq I_k'$} such that
$\mydownarrow({c_i}) \subseteq c_i'$. Hence \mbox{$[p(\vec{x})
\neck c'']_{\approx} \in I_{k+1}'$} such that $\mydownarrow(c')
\subseteq c''$ so that $I_{k+1} \sqsubseteq I_{k+1}'$ and the
result follows.
\end{proof}

\begin{proof}[Proof for theorem~\ref{theorem-approx}] \rm
The proof tactic is analogous to that used for
proposition~\ref{prop-down}.
\end{proof}

\begin{proof}[Proof for corollary~\ref{cor-semi}] \rm
Let ${\mathcal{C}}$ be a semi-cylindric constraint system and
$\rho \in uco(\wp^{\mydownarrow}(\mathcal{C}))$ be a
semi-morphism. By proposition~\ref{prop-down} it follows that
$\mydownarrow(\mathcal{F}^\mathcal{C}(P)) \sqsubseteq
\mathcal{F}^{\wp^{\mydownarrow}(\mathcal{C})}(\mydownarrow(P))$
and hence $\rho(\mydownarrow(\mathcal{F}^\mathcal{C}(P)))
\sqsubseteq
\rho(\mathcal{F}^{\wp^{\mydownarrow}(\mathcal{C})}(\mydownarrow(P)))$
and by theorem~\ref{theorem-approx}
$\rho(\mathcal{F}^{\wp^{\mydownarrow}(\mathcal{C})}(\mydownarrow(P)))
\sqsubseteq
\mathcal{F}^{\mathrm{cod}(\rho)}(\rho(\mydownarrow(P)))$ and so
the result follows.
\end{proof}


\begin{proof}[Proof for proposition~\ref{prop-coco}] \rm
Let $D_{n+1} \sqsubseteq D_{n}$ for all $n \in \mathbb{N}$. Put
$E_n = \cup \{ D_l \mid l \geq n \}$ and $E = \cap \{ E_n \mid n
\in \mathbb{N} \}$. Since $D_{n+1} \sqsubseteq D_{n}$ observe that
$E_n \equiv D_n$ for all $n \in \mathbb{N}$ and hence \linebreak
${\mathcal{D}_P^{\rho, \mathcal{C}}}(\sqcap \{ [D_n]_{\equiv} \mid n \in \mathbb{N}
\})$ = ${\mathcal{D}_P^{\rho, \mathcal{C}}}(\sqcap \{ [E_n]_{\equiv} \mid n \in
\mathbb{N} \})$ = ${\mathcal{D}_P^{\rho, \mathcal{C}}}([E]_{\equiv})$ =
$[{\mathcal{D}_P^{\rho, \mathcal{C}}}(E)]_{\equiv}$ = $[\cap \{ {\mathcal{D}_P^{\rho, \mathcal{C}}}(E_n)
\mid n \in \mathbb{N} \} ]_{\equiv}$ = $\sqcap \{
[{\mathcal{D}_P^{\rho, \mathcal{C}}}(E_n)]_{\equiv} \mid n \in \mathbb{N} \}$ =
$\sqcap \{ {\mathcal{D}_P^{\rho, \mathcal{C}}}([E_n]_{\equiv}) \mid n \in \mathbb{N}
\}$ = $\sqcap \{ {\mathcal{D}_P^{\rho, \mathcal{C}}}([D_n]_{\equiv}) \mid n \in
\mathbb{N} \}$.
\end{proof}

\begin{proof}[Proof for lemma~\ref{lemma-demand}] \rm
Proof by (double) induction. Let $\langle p(\vec{y}); c''; 1
\rangle = s_1 \rrreduce_{P}^{\star} s_n \rrreduce_{P} \lozenge$,
$s_i = \langle g_i; c_i; h_i \rangle$ and suppose
$(\mathcal{D}^{\rho, \mathcal{C}}_{P})^k(\top) = [D_k]_{\equiv}$.
The outer induction is on $k$.
\begin{description}

\item[\it base case:]
Suppose $\max(\{ \max(h_i) \mid 1 \leq i \leq n \}) \leq 1$ and
\mbox{$[p(\vec{y}) \neck e]_{\approx} \in D_1$}. Thus $s_1
\rrreduce_{P} \lozenge$ so that $s_1 \rreduce_{P} \lozenge$ and
hence there exists $p(\vec{x}') \neck C' \separate c', g' \in P$
such that
$\partial_{\vec{y}}^{\vec{x}'}(\overline{\exists}_{\vec{y}}(c''))
\not\in C'$. Then \mbox{$[p(\vec{x}') \neck e']_{\approx} \in
D_1$} where $\overline{\exists}_{\vec{z}}(d_{\vec{z}, \vec{y}}
\mywedge e) = \overline{\exists}_{\vec{z}}(d_{\vec{z}, \vec{x}'}
\mywedge e')$ and $\mathrm{var}(\vec{z}) \cap
(\mathrm{var}(\vec{y}) \cup FV(e) \cup \mathrm{var}(\vec{x}') \cup
FV(e')) = \emptyset$. Observe that $e' \subseteq
\overline{\forall}_{\vec{x}'}(C')$ and thus
$\partial_{\vec{y}}^{\vec{x}'}(\overline{\exists}_{\vec{y}}(e))$ =
$\overline{\exists}_{\vec{x}'}(e') \subseteq
\overline{\exists}_{\vec{x}'}(\overline{\forall}_{\vec{x}'}(C'))$
= $\overline{\forall}_{\vec{x}'}(C') \subseteq C'$. Hence
$\partial_{\vec{y}}^{\vec{x}'}(\overline{\exists}_{\vec{y}}(c''))
\not\in
\partial_{\vec{y}}^{\vec{x}'}(\overline{\exists}_{\vec{y}}(e))$
so that $\overline{\exists}_{\vec{y}}(c'') \not\in
\overline{\exists}_{\vec{y}}(e)$ as required.

\item[\it inductive case:]
Suppose $k = \max(\{ \max(h_i) \mid 1 \leq i \leq n \}) > 1$ and
\mbox{$[p(\vec{y}) \neck e]_{\approx} \in D_k$}. Suppose, for the
sake of a contradiction, that $\overline{\exists}_{\vec{y}}(c'')
\in \overline{\exists}_{\vec{y}}(e)$. Since $k > 1$ there exists
$w = p(\vec{x}) \neck C \separate c, p_1(\vec{x}_1), \ldots,
p_l(\vec{x}_l) \in P$, $\varphi \in Ren$ such that
$\varphi(\mathrm{CLP}(w))$ = $p(\vec{x}') \neck c',
p_1(\vec{x}_1'), \ldots, p_l(\vec{x}_l') \ll_{s_1}
\mathrm{CLP}(P)$ and $s_2 = \langle p_1(\vec{x}_1'), \ldots,
p_l(\vec{x}_l'); c_1'; 2^{l} \rangle$ and $c_1' = c'' \mywedge
d_{\vec{y}, \vec{x}'} \mywedge c'$. Suppose $\langle
p_1(\vec{x}_1'); c_1' \rangle \rreduce_{P}^{\star} \langle
\epsilon; c_2' \rangle$, \ldots, \mbox{$\langle p_m(\vec{x}_m');
c_m' \rangle \rreduce_{P}^{\star} \lozenge$}. Without loss of
generality assume $FV(\mathrm{CLP}(w)) \cap FV(c_i') = \emptyset$
for all $i \in [1, m]$. Let $\vec{v} = \vec{x} \cdot \vec{x}_1
\cdots \vec{x}_l$ and $\vec{v}' = \vec{x}' \cdot \vec{x}'_1 \cdots
\vec{x}'_l$. Let $g_i' \in \mathcal{C}$ such that $\langle
p_i(\vec{x}_i'); 1 \rangle \reduce_{P}^{\star} \langle \epsilon;
g_i' \rangle$ and $c_{i+1}' = c_{i}' \mywedge g_{i}'$ for all $i
\in [1, m)$. For all $i \in [1, m)$, put $g_i =
\partial_{\vec{x}_i'}^{\vec{x}_i}(g_i')$. Put $c_1 =
\partial_{\vec{v}'}^{\vec{v}}(c_1')$ and for all $i \in [2, m]$, put
$c_i = \partial_{\vec{x}_i'}^{\vec{x}_i}(c_i')$. Then $c_{i+1} =
c_{i} \mywedge g_{i}$ for all $i \in [1, m)$. Let
$\mathcal{O}^{\mathcal{C}}(P) = [F]_{\equiv}$. By
proposition~\ref{prop-trivial}, \mbox{$[p_i(\vec{x}_i) \neck
g_i]_{\approx}$} = \mbox{$[p_i(\vec{x}_i') \neck g_i']_{\approx}
\in F$} for all $i \in [1, m)$. By theorem~\ref{theorem-equiv},
$\mathcal{O}^{\mathcal{C}}(P) = \mathcal{F}^{\mathcal{C}}(P)$ and
by corollary~\ref{cor-semi},
$\rho(\mydownarrow(\mathcal{F}^{\mathcal{C}}(P))) \sqsubseteq
\mathcal{F}^{\mathrm{cod}(\rho)}(\rho(\mydownarrow(P)))$. Thus for
$i \in [1, m)$ there exists $[p_i(\vec{x}_i) \neck f_i]_{\approx}
\in F$ such that $\rho(\mydownarrow(g_i)) \subseteq f_i$. Put $f_i
= 0$ for all $i \in [m, l]$ to ensure $[p_i(\vec{x}_i) \neck
f_i]_{\approx} \in F$ for all $i \in [m, l]$. Let $[p_i(\vec{x}_i)
\neck d_i]_{\approx} \in D_k$ for all $i \in [1, l]$. Finally put
$e_{n+1} = \mathcal{C}$, $e_{i} = d_i \cap (f_i \rightarrow'
e_{i+1})$ for all \mbox{$i \in [1, l]$} and $e_0 = C \cap
(\rho(\mydownarrow(c)) \rightarrow' e_1)$. The inner induction is
on $i$ and is used to show $\rho(\mydownarrow(c_{i})) \subseteq
e_i$ for all $i \in [1, m]$.
\begin{description}

\item[\it base case:] Now $c_1 =
\partial_{\vec{v}'}^{\vec{v}}(c_1') \mymodels \partial_{\vec{y}}^{\vec{x}}(
\overline{\exists}_{\vec{y}}(c''))
\in C$. Thus
$\rho(\mydownarrow(c_1)) \subseteq C$. Furthermore, $c_1 =
\partial_{\vec{v}'}^{\vec{v}}(c_1') \mymodels \partial_{\vec{v}'}^{\vec{v}}(c') = c$.
Thus $\rho(\mydownarrow(c_1)) \subseteq \rho(\mydownarrow(c))$.
Moreover, $c_1 \mymodels \partial_{\vec{y}}^{\vec{x}}(\overline{\exists}_{\vec{y}}(c''))
\in
\partial_{\vec{y}}^{\vec{x}}(\overline{\exists}_{\vec{y}}(e)) \subseteq
\overline{\forall}_{\vec{x}}(e_0) \subseteq e_0$.
Thus $\rho(\mydownarrow(c_1)) \subseteq e_0$.
However, $e_0 = C \cap (\rho(\mydownarrow(c))
\rightarrow' e_1)$. Thus $\rho(\mydownarrow(c_1)) \subseteq C \cap
(\rho(\mydownarrow(c)) \rightarrow' e_1)$ and
$\rho(\mydownarrow(c_1)) \subseteq \rho(\mydownarrow(c_1)) \cap C \cap
(\rho(\mydownarrow(c)) \rightarrow' e_1)$ =
$\rho(\mydownarrow(c_1)) \cap
(\rho(\mydownarrow(c)) \rightarrow' e_1)$ =
$\rho(\mydownarrow(c_1)) \cap \rho(\mydownarrow(c)) \cap
(\rho(\mydownarrow(c)) \rightarrow' e_1)$ =
$\rho(\mydownarrow(c_1)) \cap \rho(\mydownarrow(c)) \cap e_1$
=
$\rho(\mydownarrow(c_1)) \cap e_1$.
Therefore
$\rho(\mydownarrow(c_1)) \subseteq e_1$ as required.

\item[\it inductive case:] Suppose $\rho(\mydownarrow(c_{i}))
\subseteq \varphi(e_i)$. Now $\rho(\mydownarrow(c_{i+1}))$ =
$\rho(\mydownarrow(c_{i} \mywedge g_i))$ \linebreak \mbox{$\subseteq
\rho(\mydownarrow(c_{i})) \cap \rho(\mydownarrow(g_{i}))$}
\mbox{$\subseteq e_i \cap \rho(\mydownarrow(g_{i}))$}
\mbox{$\subseteq e_i \cap \rho(\mydownarrow(g_{i}))$}
\mbox{$\subseteq e_i \cap f_i$} \mbox{$\subseteq (f_i \rightarrow'
e_{i+1}) \cap f_i$} = $e_{i+1}$. Therefore
$\rho(\mydownarrow(c_{i+1})) \subseteq e_{i+1}$ as required.

\end{description}

Thus $\rho(\mydownarrow(c_m)) \subseteq e_m \subseteq d_m$ so that
$c_m \in d_m$. Let $d_m' =
\partial_{\vec{x}_m}^{\vec{x}_m'}(\overline{\exists}_{\vec{x}_m}(d_m))$
and observe that $[p_m(\vec{x}'_m) \neck d'_m]_{\approx}$ =
$[p_m(\vec{x}_m) \neck d_m]_{\approx} \in D_k$. Put $c_m'' =
\partial_{\vec{x}_m}^{\vec{x}_m'}(\overline{\exists}_{\vec{x}_m}(c_m))$
so that $c_m' \mymodels c_m'' \in d_m'$. By the inductive hypothesis
\mbox{$\langle p_m(\vec{x}'_m); c'_m \rangle
\not\rreduce_{P}^{\star} \lozenge$} which is a contradiction and hence
$\overline{\exists}_{\vec{y}}(c'') \not\in \overline{\exists}_{\vec{y}}(e)$
as required.
\end{description}

\noindent The result follows.
\end{proof}

\begin{proof}[Proof for theorem~\ref{theorem-demand}] \rm
Let $\mathcal{D}^{\rho, \mathcal{C}}(P) = [D]_{\equiv}$,
$[p(\vec{y}) \neck e]_{\approx} \in D$ and $c'' \in
\overline{\exists}_{\vec{y}}(e)$. Thus
$\overline{\exists}_{\vec{y}}(c'') \in
\overline{\exists}_{\vec{y}}(\overline{\exists}_{\vec{y}}(e))$ =
$\overline{\exists}_{\vec{y}}(e)$. Suppose, for the sake of a
contradiction, that $\langle p(\vec{y}); c''; 1 \rangle = s_1
\rrreduce_{P}^{\star} s_n \rrreduce_{P} \lozenge$ where $s_i =
\langle g_i; c_i; h_i \rangle$. Let $k = max(\{ \max(h_i) \mid 1
\leq i \leq n \})$. Suppose $(\mathcal{D}^{\rho,
\mathcal{C}}_{P})^k(\top) = [D_k]_{\equiv}$. Since $D \sqsubseteq
D_k$ and by lemma~\ref{lemma-demand} there exists $[p(\vec{y})
\neck e']_{\approx} \in D_k$ such that
$\overline{\exists}_{\vec{y}}(c'') \not\in
\overline{\exists}_{\vec{y}}(e')$. Since
$\overline{\exists}_{\vec{y}}(e) \subseteq
\overline{\exists}_{\vec{y}}(e')$ it follows that
$\overline{\exists}_{\vec{y}}(c'') \not\in
\overline{\exists}_{\vec{y}}(e)$ which is a contradiction. The
result follows.
\end{proof}

\end{document}